\documentclass[%
 reprint,
 amsmath,amssymb,
 aps,
]{revtex4-1}

\RequirePackage{lineno} 

\usepackage{subfigure}
\usepackage{graphicx}
\usepackage{dcolumn}
\usepackage{bm}
\usepackage{float}

\begin{document}

\preprint{APS/123-QED}

\title{QCD results from ATLAS}

\author{Nicola Orlando on behalf of the ATLAS Collaboration}
\affiliation{%
 INFN sez. Lecce and Dipartimento di Matematica e Fisica ``Ennio De Giorgi'', Universit\`a del Salento\\
}%



\begin{abstract}
The most recent QCD measurements performed in ATLAS are reviewed; 
the results summarized here are based on data collected with the ATLAS detector during the 2010 and 2011 data taking 
in proton-proton collisions at center of mass energy $\mathrm{\sqrt s = 7}$ TeV at the Large Hadron Collider, 
corresponding approximately to an integrated luminosity of 36~pb$^{-1}$ and 4.6~fb$^{-1}$ respectively. 

\end{abstract}

\pacs{Valid PACS appear here}
\maketitle


\section{Introduction}

High energy proton-proton collisions at the Large Hadron Collider (LHC) offer a natural ground for studying the strong 
interaction by testing its perturbative description 
provided by the Quantum-Chromo-Dynamic (QCD) sector of the Standard Model (SM) as well as by validating the modeling of the non-perturbative phenomena 
in current Monte Carlo (MC) event generators.

Two particularly important
areas of study are the production of jets and of vector gauge bosons;
these processes lead to backgrounds to Higgs boson analysis
as well as to Beyond Standard Model (BSM) searches. Thus, detailed measurements of jets and vector bosons are mandatory. 

Cross section measurements performed in ATLAS are corrected for detector effects to the particle level in a 
fiducial volume chosen as close as possible to the detector acceptance. Particle level distributions measured in data are compared to the predictions of Monte Carlo 
event generators, which include models for parton hadronization as well as multiple-parton-interactions (MPI), and parton level predictions corrected by 
non-perturbative effects at particle level. 
 
This paper is organized in four main sections: 
selected measurements of jets are described in Sec.~\ref{sec:Jets}; the most recent analysis of isolated photon production will be summarized 
in Sec.~\ref{sec:Photons}; in Sec.~\ref{sec:WZ} and Sec.~\ref{sec:WZJets} measurements of 
inclusive gauge boson and gauge boson production in association with jets will be reported. 
 

\section{QCD with jets\label{sec:Jets}}

\begin{figure}[ht]
  \subfigure{\includegraphics[width=0.85\linewidth]{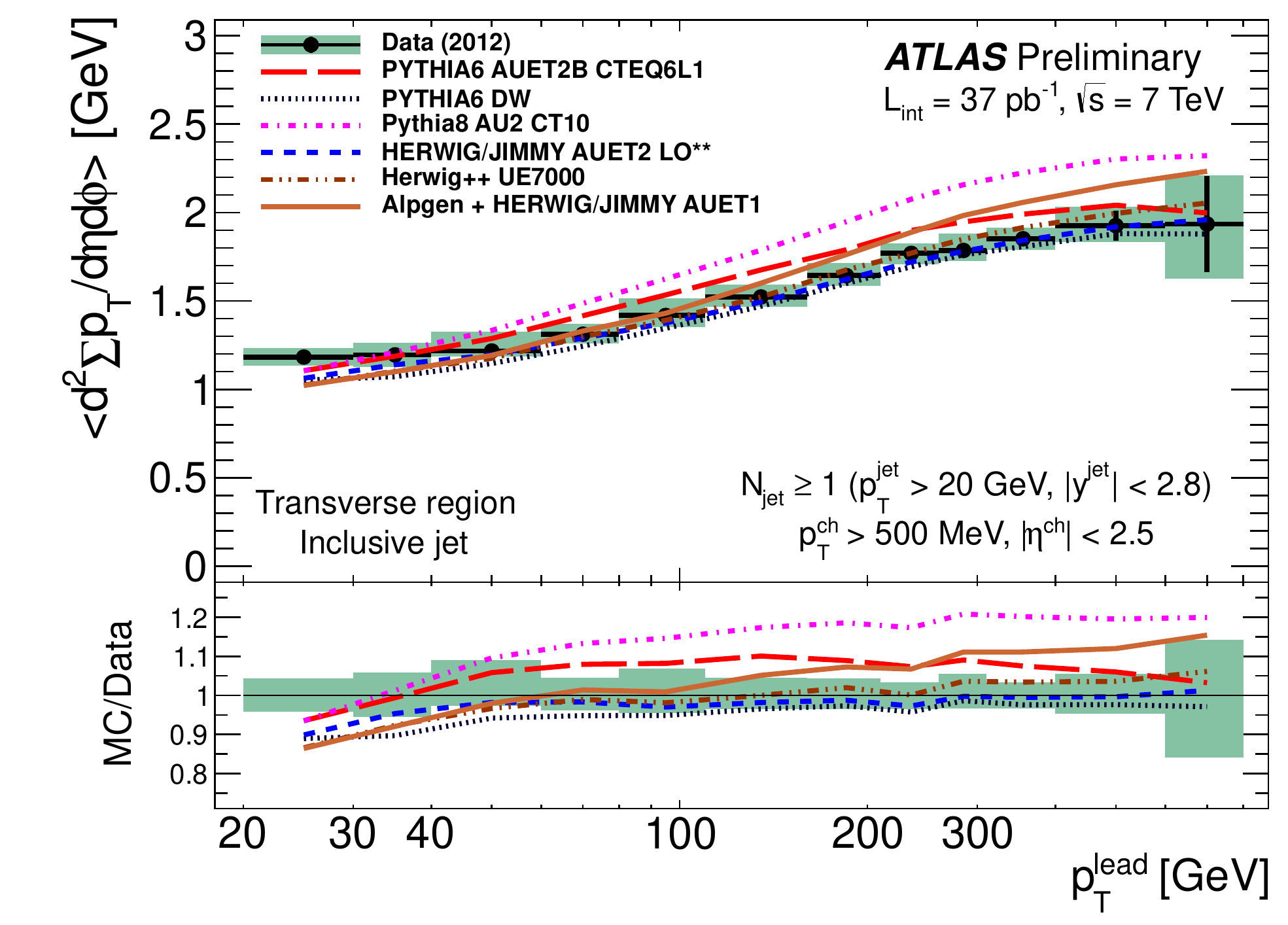}}
  \subfigure{\includegraphics[width=0.85\linewidth]{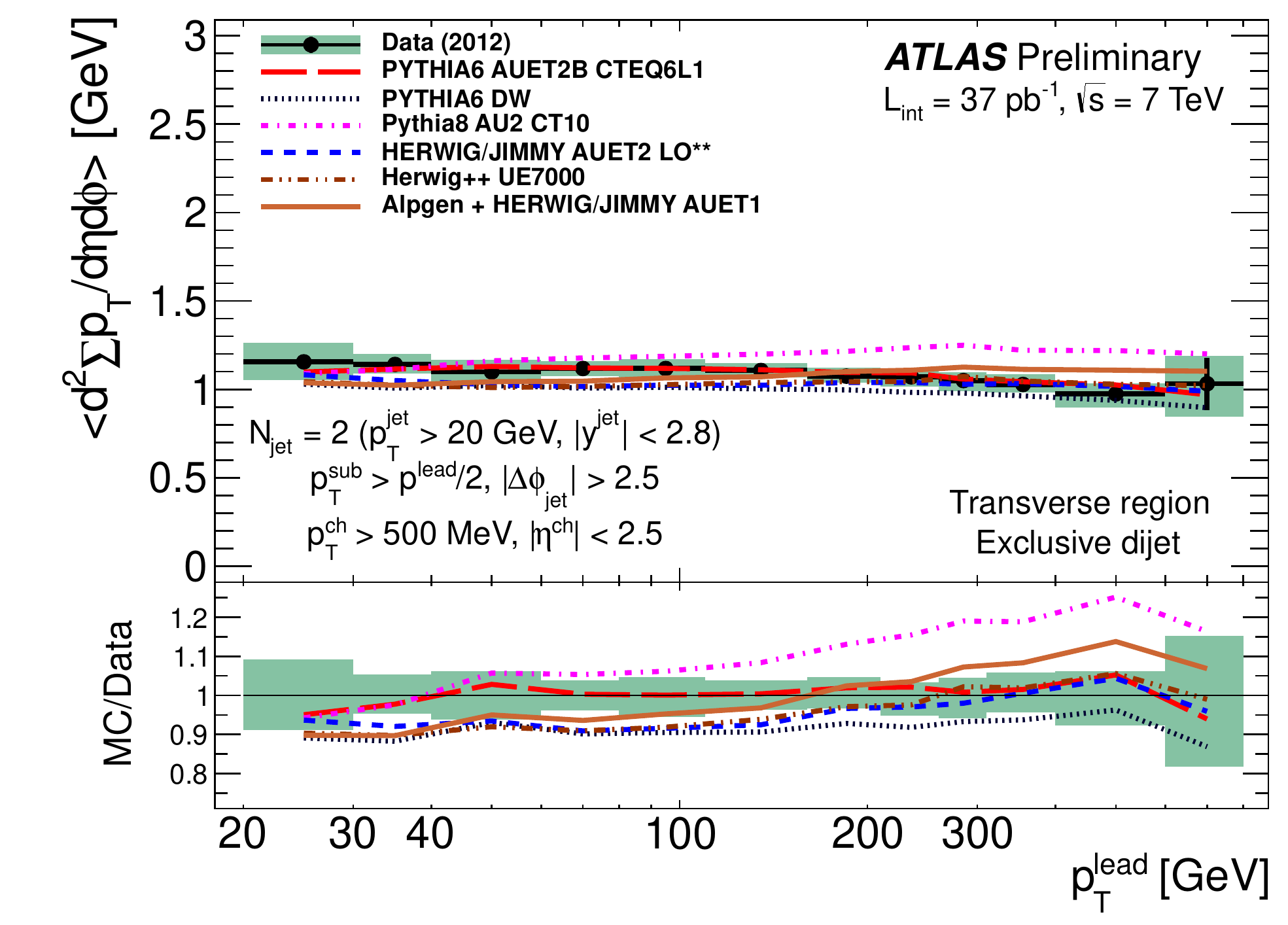}}
  \caption{Profiles of charged particle $\mathrm{\sum p_T}$ as a function of p$\mathrm{_{T}^{lead}}$~\cite{ATLAS-CONF-2012-164} for the inclusive jet event selection (top) and exclusive dijet event selection (bottom).
    The green bands around the data points represent the statistical and systematic uncertainties added in quadrature.
    The data are compared to several MC models based on the description of the UE given by 
    {\sc{Pythia8}}~\cite{Sjostrand:2007gs}, {\sc{Herwig++}}, 
    {\sc{Pythia6}} and {\sc{Herwig+Jimmy}}.\label{fig:UE}}
\end{figure}

Hadronic jets are copiously produced at the LHC allowing multi-differential and exclusive measurements with relatively small integrated luminosity. 
Moreover, the signature provided by high-p$\mathrm{_T}$ jets can be used as reference to study beam-beam remnants and initial/final state radiation (ISR/FSR) and MPI, 
together referred to as the 
underlying event (UE). 

The underlying event characteristics can be studied by measuring the properties of particles accompanying high-p$\mathrm{_T}$ jet~\cite{ATLAS-CONF-2012-164}. 
The selected events are classified as events with at least one selected jet and events with exactly two jets, 
allowing an increase in the relative sensitivity to different components of the UE. 
In Fig.~\ref{fig:UE} the distribution of the variable: 
\begin{equation}
\mathrm{ \sum p_T \equiv \left \langle \frac{d^{2}\sum_{tracks} p_{T}}{d\eta d\phi} \right \rangle \, , }
\end{equation} 
is shown as a function of the transverse momentum of the leading jet, p$\mathrm{_{T}^{lead}}$. 
A slow rise with p$\mathrm{_{T}^{lead}}$ of the UE activity is observed, as expected, for the distribution obtained with the inclusive jet selection (Fig.~\ref{fig:UE}, top).
On the other hand, selecting exclusive dijet events, $\mathrm{\sum p_T}$ exhibits less sensitivity to high order QCD effects, 
so the distribution shows a flat behavior as a function of p$\mathrm{_{T}^{lead}}$ as can be observed in Fig.~\ref{fig:UE} (top).
\begin{figure}[H]
  \subfigure{\includegraphics[width=0.95\linewidth]{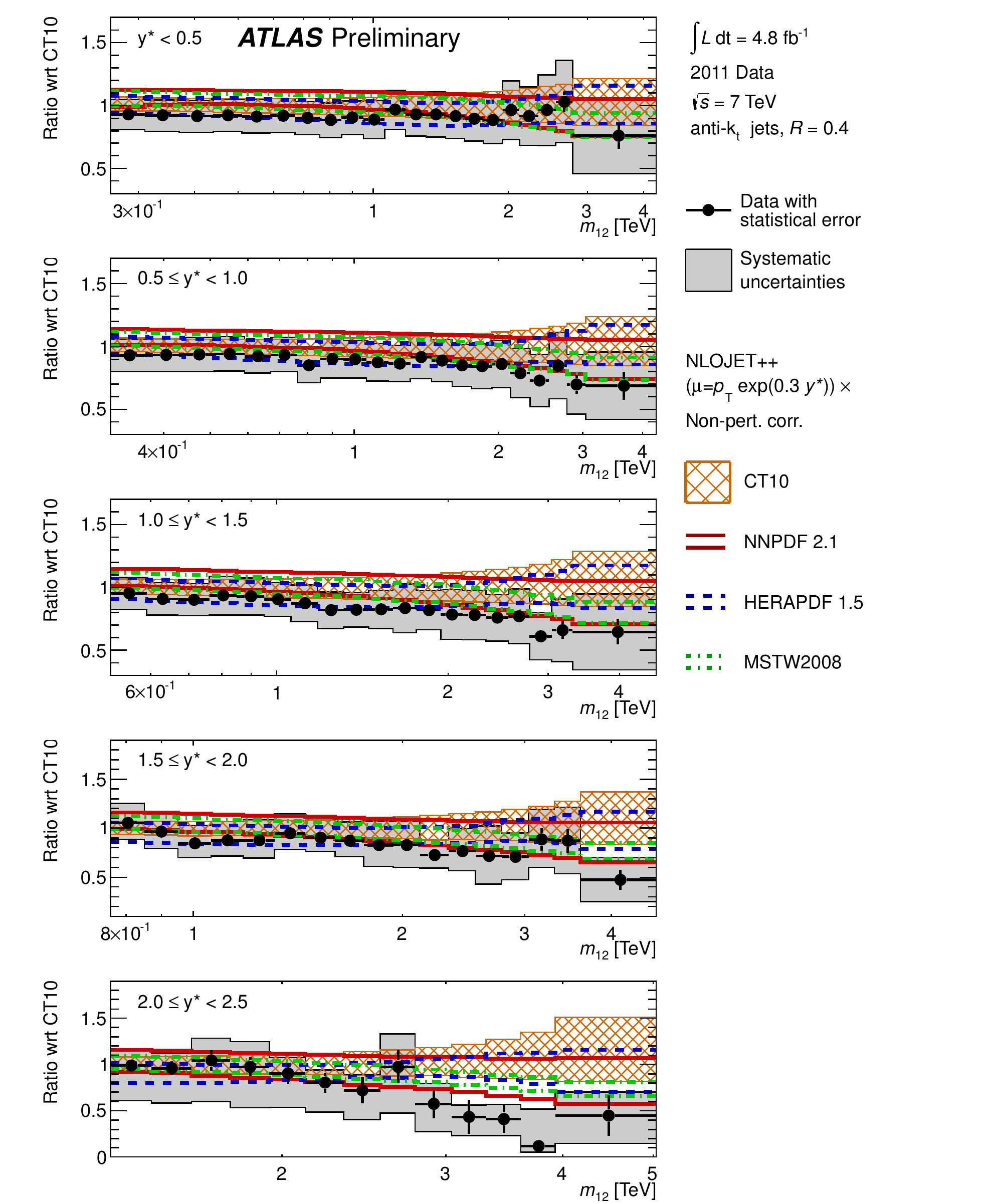}}
  \caption{Ratios of dijet double-differential cross section~\cite{ATLAS-CONF-2012-021} 
    to the theoretical prediction obtained using {\sc{Nlojet++}} with the {\sc{Ct10}} PDF set. 
    Results are shown for jets identified using the anti-k$\mathrm{_t}$ algorithm with radius parameter R$=0.4$. 
    The total systematic uncertainties on the measurement are indicated by the dark-shaded bands.
    The data are compared to various predictions obtained with {\sc{Nlojet++}} calculation using the PDF sets 
    {\sc{Ct10}}, {\sc{Mstw2008}}, {\sc{Nnpdf2.1}} and {\sc{Herapdf1.5}}.\label{fig:Jet}} 
\end{figure}
\begin{figure}[th]
  \subfigure{\includegraphics[width=0.75\linewidth]{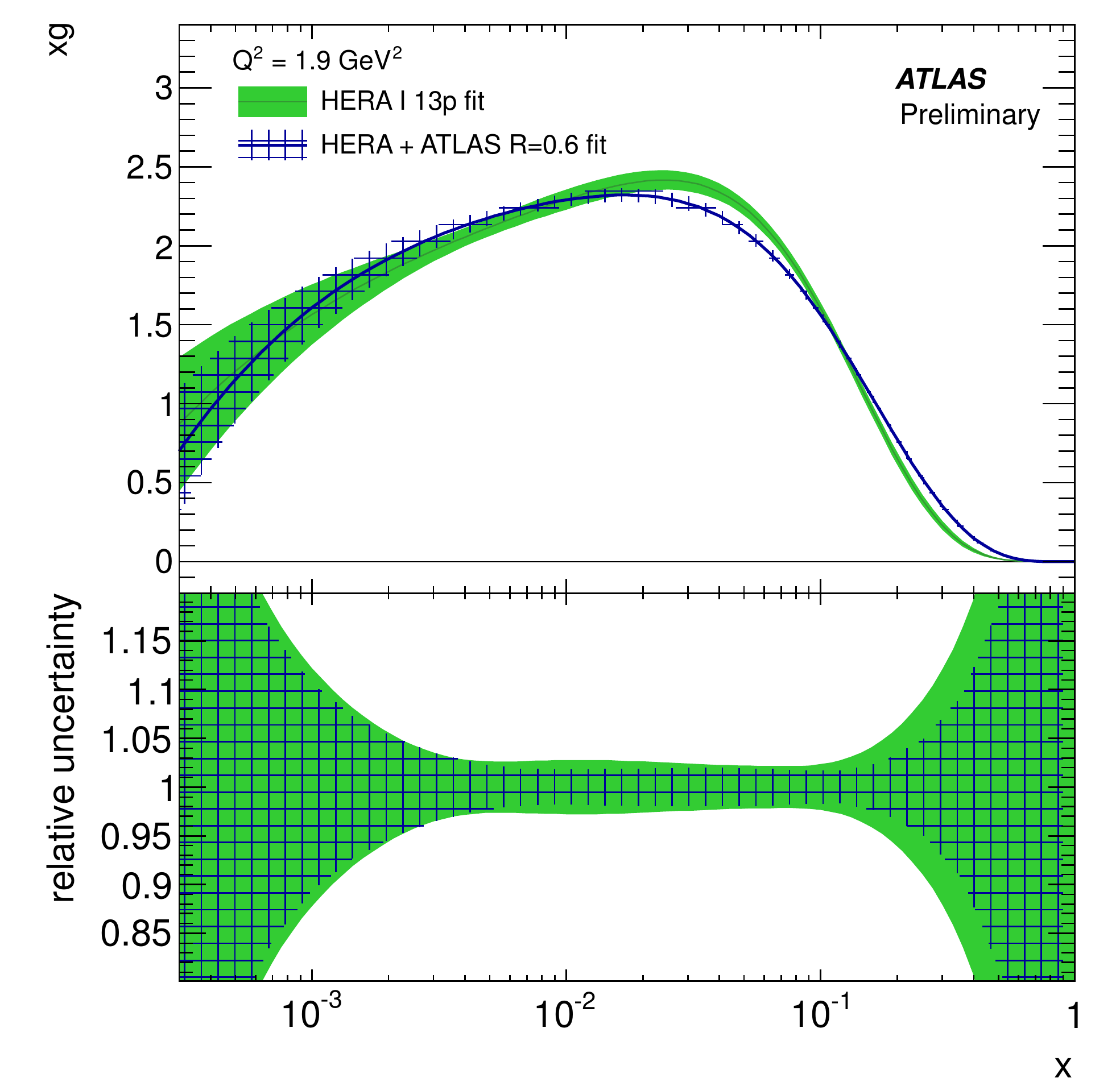}}
  \caption{Gluon density xg and its relative uncertainty as a function of x for $\mathrm{Q^2=1.9}$ GeV$\mathrm{^2}$~\cite{ATLAS-CONF-2012-128}. 
    The green band indicates a fit to HERA data only. 
    The blue band shows a fit to HERA$+$ATLAS jet data for jets identified using the anti-k$\mathrm{_t}$ algorithm with radius parameter R$=0.6$ size. 
    For each fit the uncertainty of the PDF is centered on unity. \label{fig:GuonPdf}}
\end{figure}
\begin{figure*}[tbp]
    \begin{tabular}{ccc}
      \includegraphics[width=0.32\textwidth]{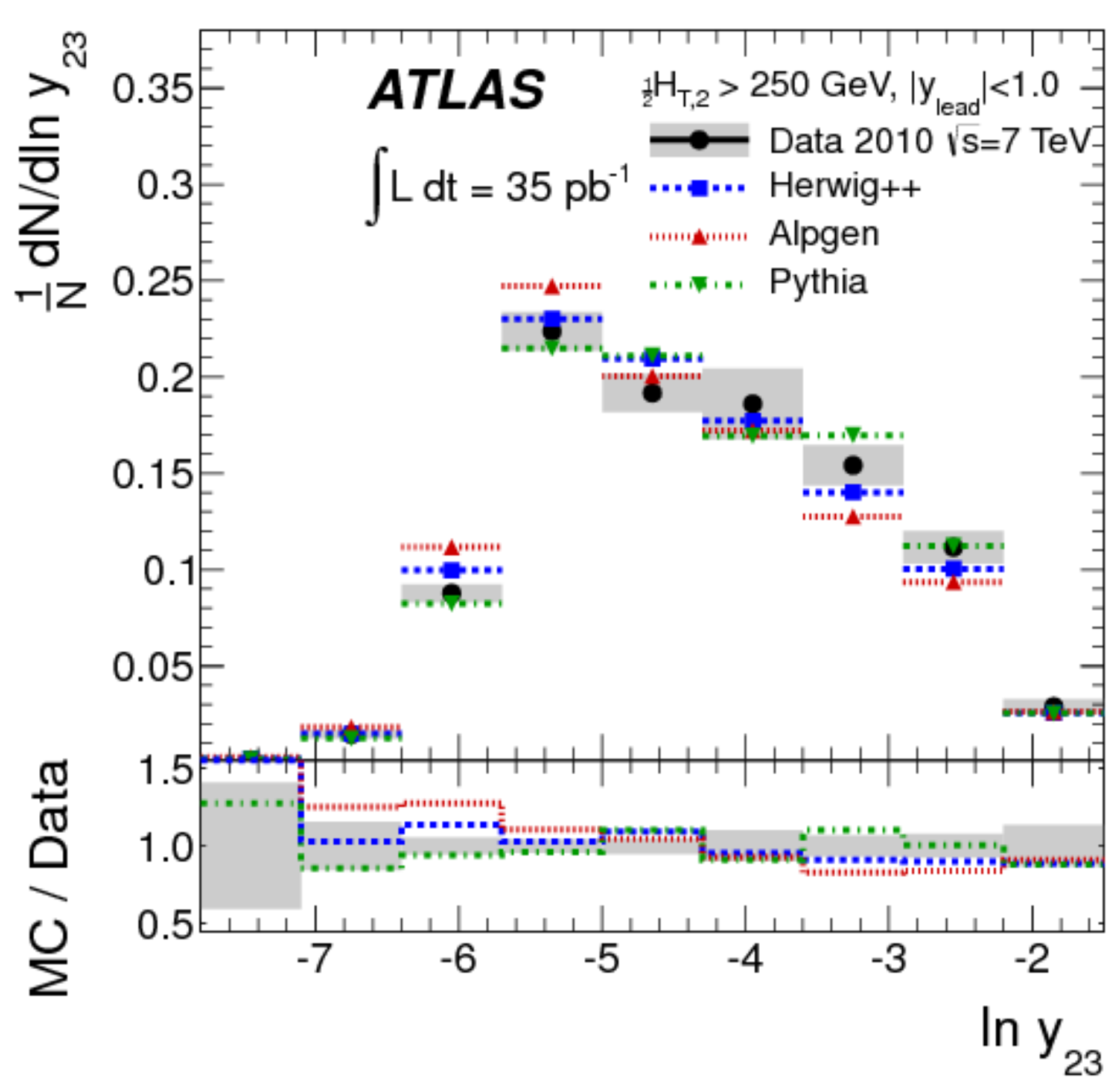} &
      \includegraphics[width=0.32\textwidth]{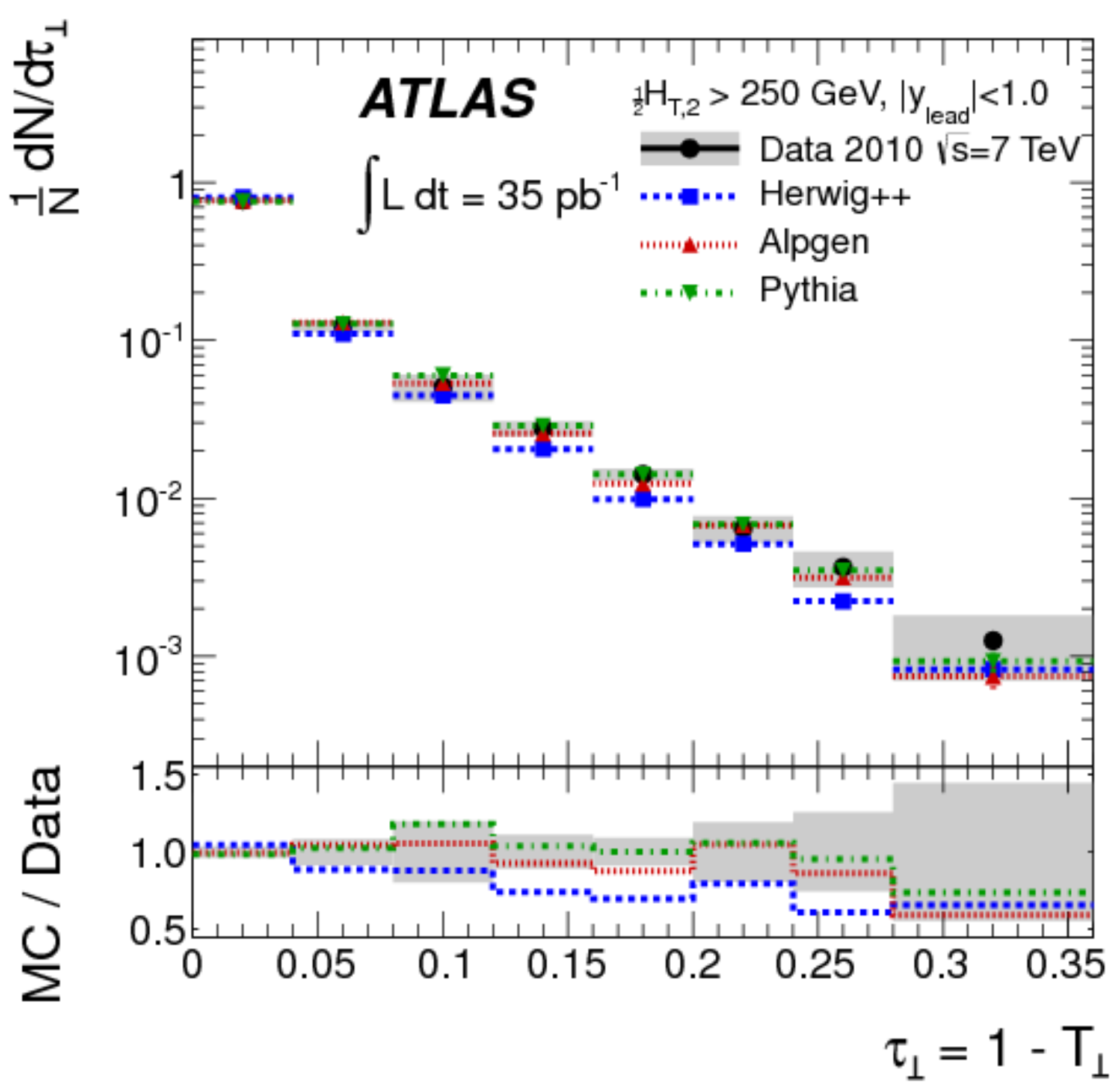} &
      \includegraphics[width=0.32\textwidth]{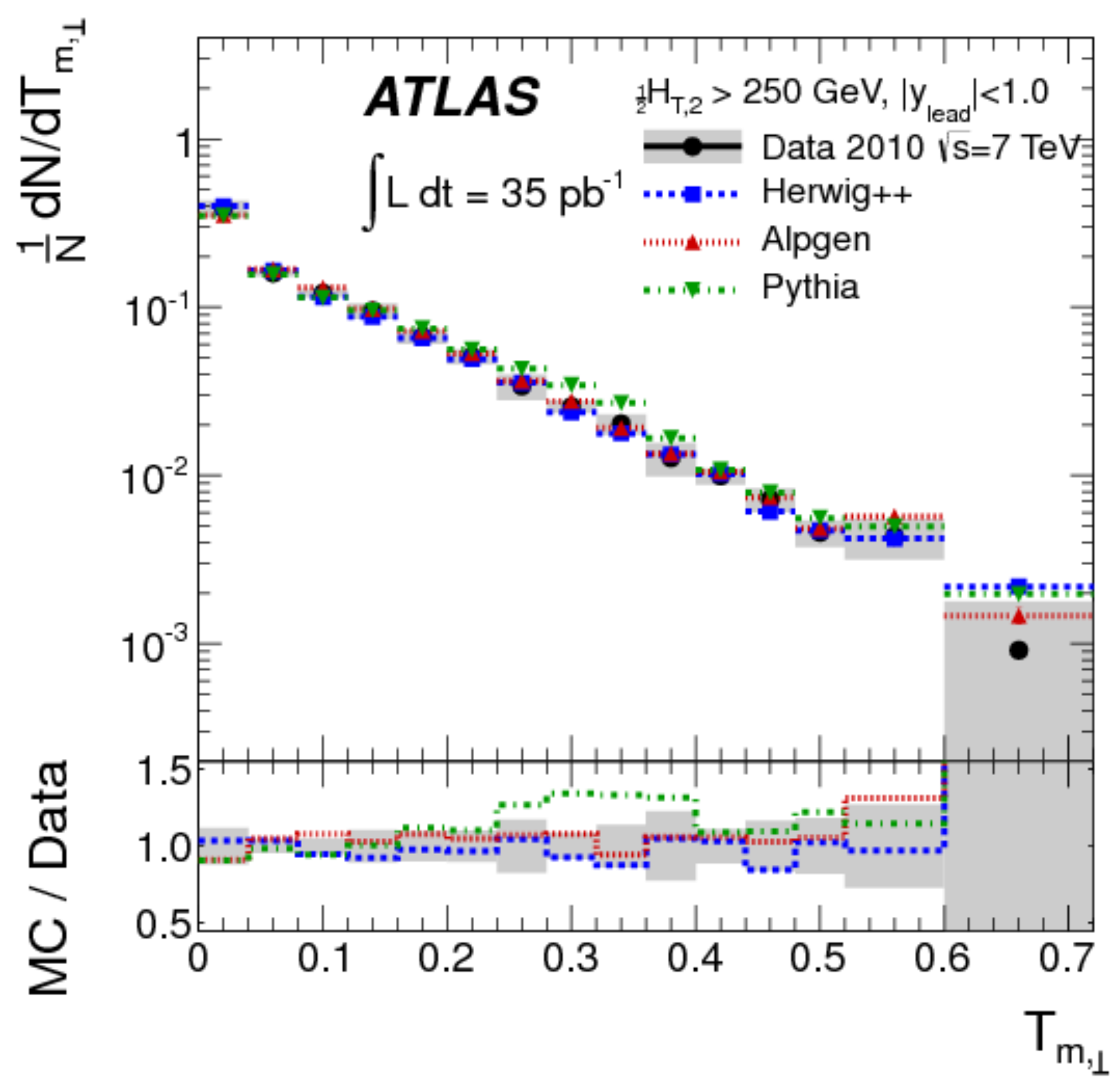} \\
    \end{tabular}
  \caption{Distributions of the third-jet resolution parameter ($\mathrm{\ln(y_{23})}$), the transverse thrust ($\tau_{\perp}$) 
    and the minor component of the transverse thrust ($\mathrm{ T_{m, \perp}}$)~\cite{Aad:2012np}. 
    The measurement is compared to LO MC generators {\sc{Alpgen}}, {\sc{Pythia6}} and {\sc{Herwig++}}.\label{fig:EventShapes}}
\end{figure*}
\begin{figure*}[tbp]
  \begin{tabular}{ccc}
    \includegraphics[width=0.32\textwidth]{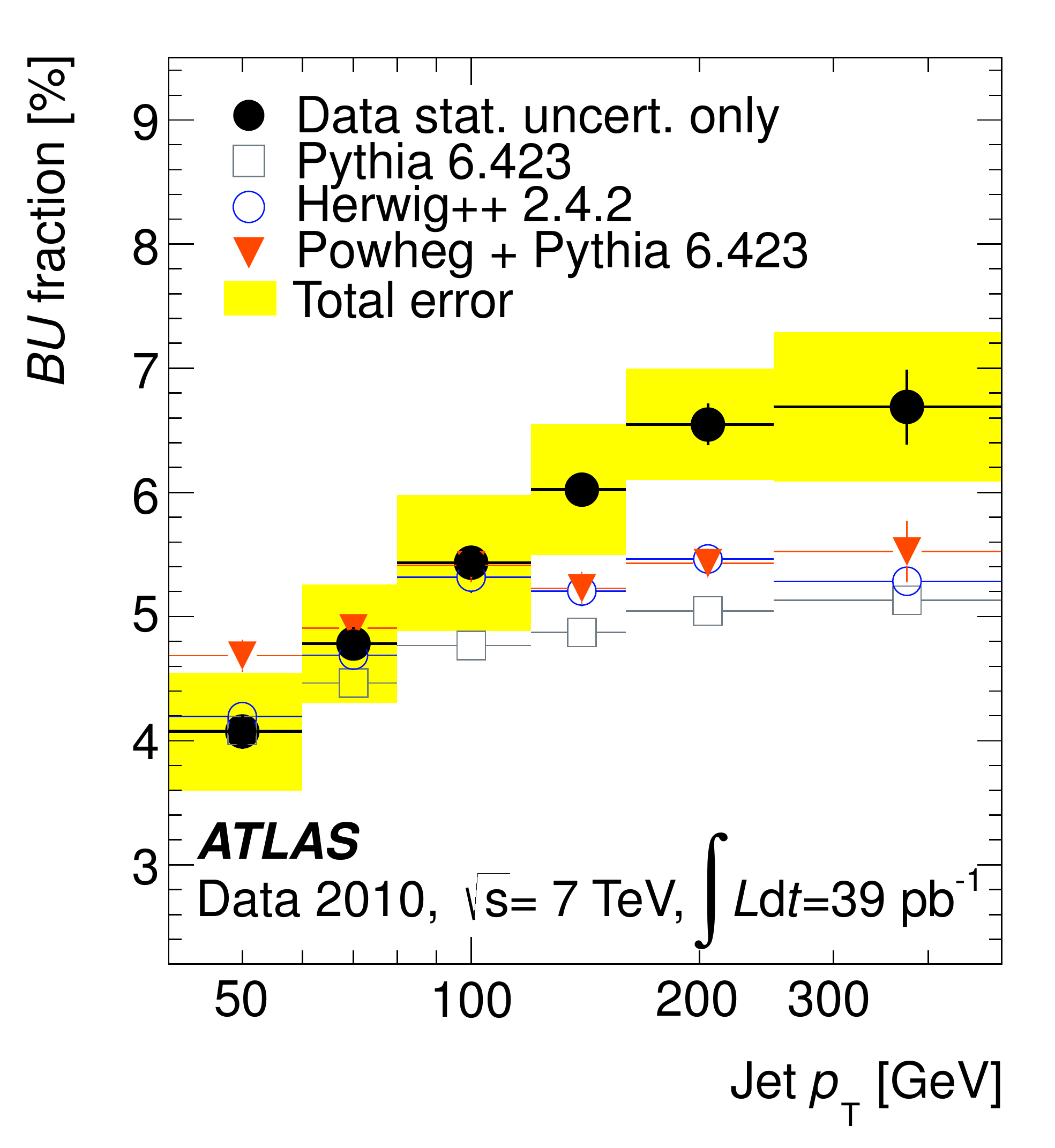} &
    \includegraphics[width=0.32\textwidth]{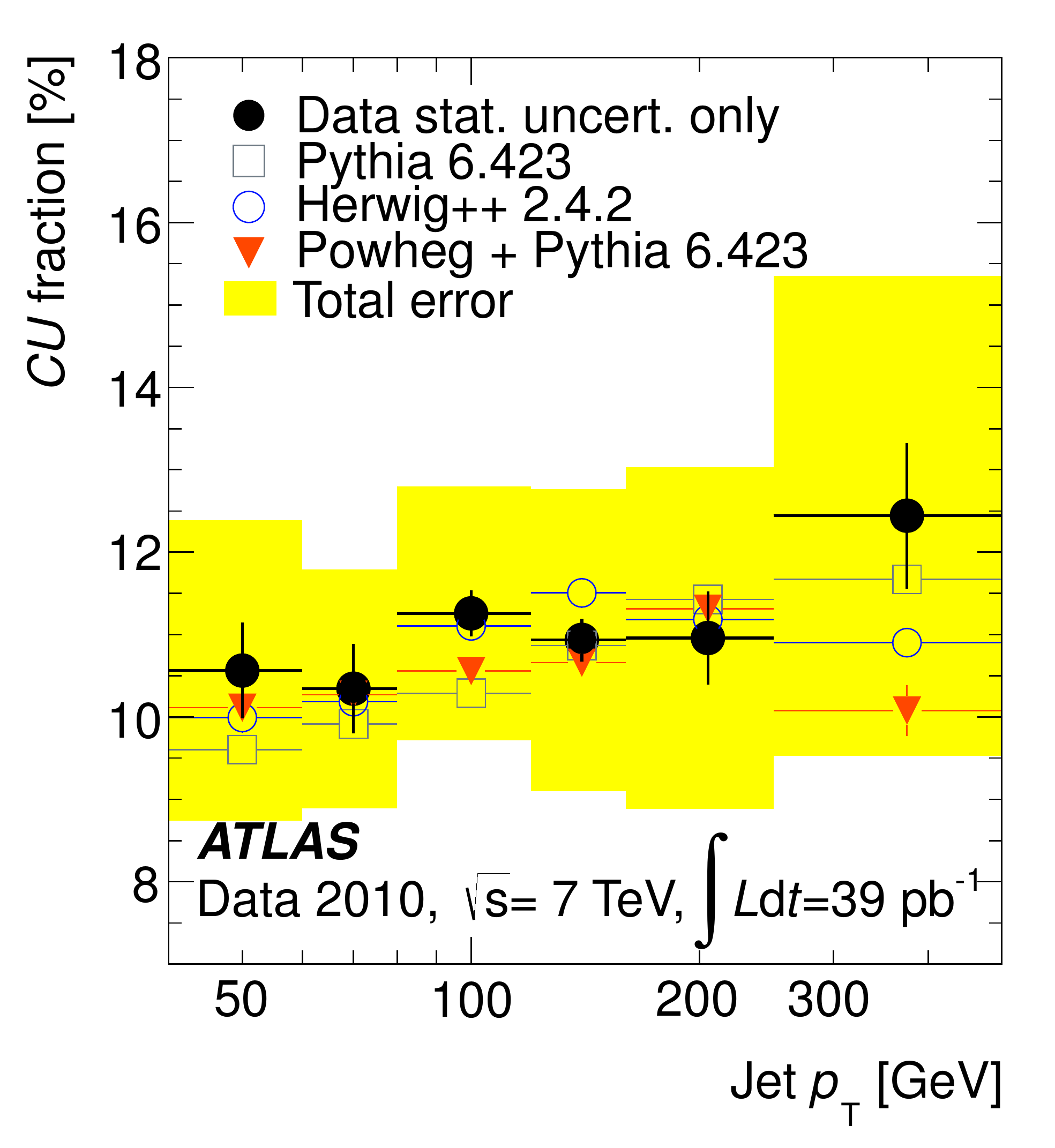} &
    \includegraphics[width=0.32\textwidth]{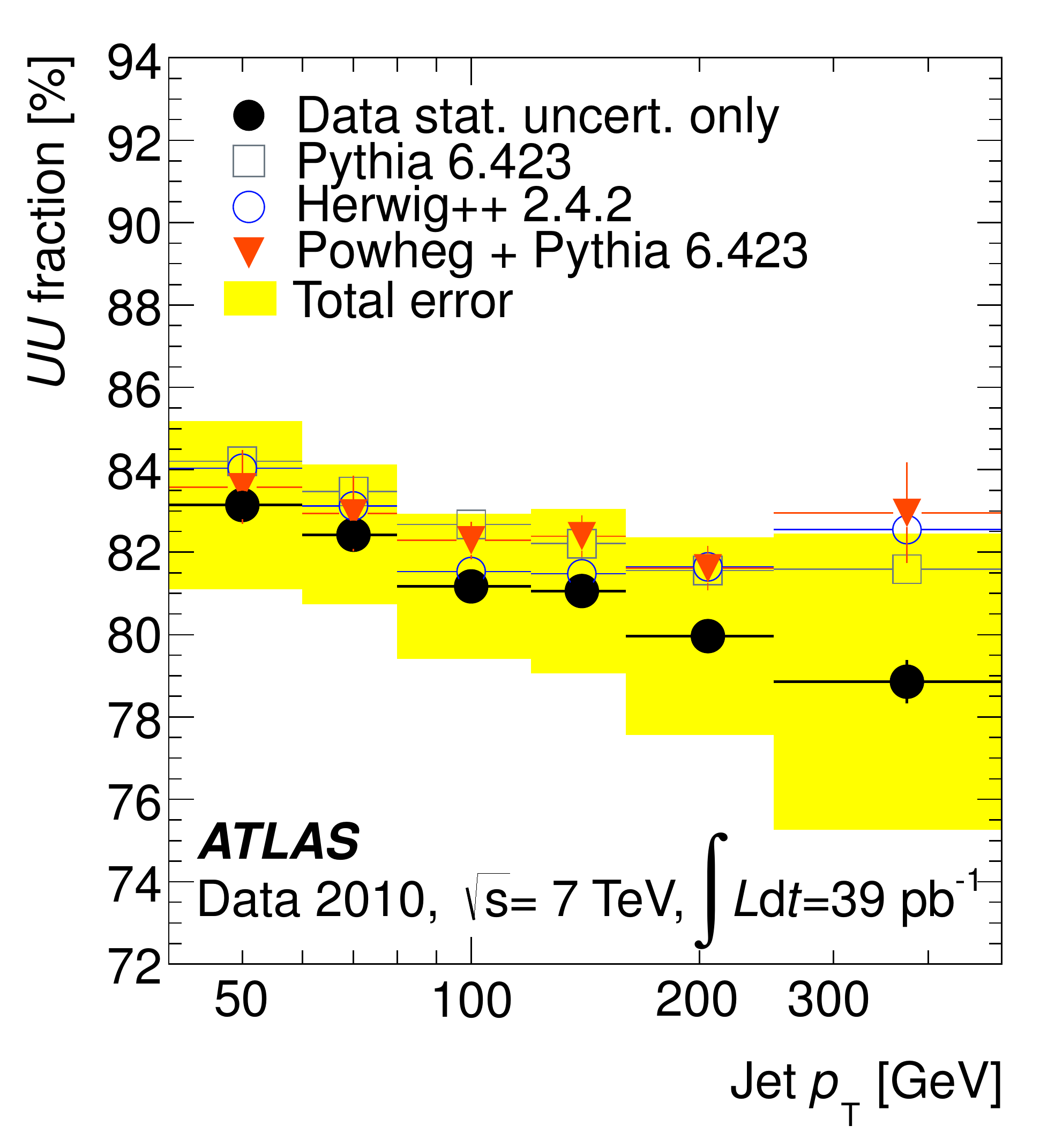} \\
  \end{tabular}
  \caption{The dijet flavor fractions~\cite{Aad:2012ma} for each leading jet $\mathrm{p_T}$ bin (black points); 
    the predictions of {\sc{Pythia6.423}} (squares), {\sc{Herwig++ 2.4.2}} (circles) and {\sc{Powheg+Pythia 6.423}} (filled triangles) are overlaid. 
    The error bars on the data points show statistical uncertainties only, whereas the total uncertainties are represented by the shaded band.\label{fig:FlavorComp}}
\end{figure*}

Several measurement of jet cross section heve been performed with the ATLAS detector using the 2010 data~\cite{Aad:2011fc,Aad:2011jz}. 
A double differential dijet cross section was recently measured using the full 2011 dataset~\cite{ATLAS-CONF-2012-021}; 
the cross section is presented as a function of the invariant mass of the two leading jets,  
binned in half the absolute rapidity difference between the two leading jets 
\begin{equation}
\mathrm{ y^*=\frac{|y_{lead}-y_{sub\;lead}|}{2} \,, }
\end{equation} 
reaching values of the invariant mass above $4$ TeV. 
The measured cross section is shown in Fig.~\ref{fig:Jet}; the data are normalized to the next-to-leading order (NLO) QCD 
calculation of {\sc{Nlojet++}}~\cite{Nagy:2003tz}, obtained with the {\sc{Ct10}}~\cite{Lai:2010vv} parton distribution function (PDF) set, 
and are compared to several NLO QCD predictions obtained using the {\sc{Mstw2008}}~\cite{Martin:2009iq}, {\sc{Nnpdf2.1}}~\cite{Ball:2010de} and {\sc{Herapdf1.5}} PDF sets.
The measurement is affected by sizable experimental systematic uncertainties, ranging from $10\%$ to $60\%$, 
dominated by the uncertainty on the jet-energy-scale (JES) determination.
However, the jet cross section measurements can be optimized, reducing experimental as well as theoretical uncertainties, 
by defining suitable observables.
In order to obtain an observable almost unaffected by the JES uncertainty, the jet cross section 
ratio at $2.76$ and $7$ TeV, $\mathrm{ \rho(y,p_T) }$:
\begin{equation}
\mathrm{ \rho(y,p_T)=\frac{\sigma^{(2.76\; TeV)}(y,p_T)}{\sigma^{(7\; TeV)} (y,p_T)} }
\end{equation} 
was measured exploiting a data sample with integrated luminosity of 20~pb$^{-1}$ collected at $\mathrm{ \sqrt s =2.76 }$ TeV at the begin of the 2011 data taking.
In the cross section ratio at two different energies several large experimental systematic uncertainties are correlated;
in particular, the uncertainty due to the JES is greatly reduced in the cross section ratio measured as a function of p$\mathrm{ _{T}^{lead} }$.
The data exhibit typically a better precision~\cite{ATLAS-CONF-2012-128} than the   
the theory predictions which suffer from uncertainties due to missing high order terms in the perturbative expansion 
and errors on parton distribution functions (PDFs) which are the largest source of theory
uncertainties in the high p$\mathrm{ _{T}}$ region 
(jet p$\mathrm{_{T}\gtrsim 100 }$ GeV). 

Given the experimental precision and the sensitivity 
of the prediction to the PDFs, the measurement of the cross section ratio was used jointly with the HERA I data~\cite{Aaron:2009aa} to perform a PDF fit, 
using the framework described in~\cite{Carli:2010rw}, in order to determine the gluon parton distribution function, $\mathrm{ g(x) }$. 
The result is compared in Fig.~\ref{fig:GuonPdf} to a fit to the HERA I data only; the inclusion of the ATLAS data in the PDF fit improves the 
uncertainty on the gluon PDF, in the high Bjorken-x region, and lead also to a harder spectrum for $\mathrm{ g(x) }$.

A characterization of event topology in multijet events can be based on the measurements of standard event shape variables; 
several event shape distributions have been reported in Ref.~\cite{Aad:2012np}, 
including the third jet resolution ($\mathrm{ \ln\, y_{23} }$), event thrust ($\mathrm{\tau_{\perp}}$) and thrust minor ($\mathrm{T_{m,\perp}}$) which are 
shown in Fig.~\ref{fig:EventShapes}. The data are reasonably well described by leading-order (LO) Monte Carlo event generators 
{\sc{Alpgen}}~\cite{Mangano:2002ea}, {\sc{Pythia6}}~\cite{Sjostrand:2006za} and {\sc{Herwig++}}~\cite{Bahr:2008pv}.

More information about the relative importance of the various jet production sub-channels can be obtained 
with a flavor classification of the high p$_\mathrm{T}$ 
jets.
The extraction of the flavor components in exclusive dijet events was presented in~\cite{Aad:2012ma}. 
The flavor assignments is based on geometrical matching of the jets with hadrons; the possible flavor assignment of the dijet system are: 
bottom-bottom (BB), charm-charm (CC), light-light (UU), bottom-charm (BC), bottom-light (BU), charm-light (CU). 
These measurements allow a probe of the relative contribution of ``quark pair creation'', ``heavy quark excitation'' and ``gluon splitting'' in heavy flavor jet production.
The BU, CU and UU fractions are shown in Fig.~\ref{fig:FlavorComp}
as a function of the leading jet p$\mathrm{_T}$; the data are compared to the predictions obtained with LO MC generators, {\sc{Pythia6.423}} and {\sc{Herwig++ 2.4.2}},  
as well as a NLO QCD calculation of {\sc{Powheg+Pythia 6.423}}~\cite{Alioli:2010xd}.
Good agreement is observed between data and QCD predictions with the remarkable exception of the BU 
fraction where a departure of the theory predictions from the data is observed 
in the high p$_T$ region, namely for leading jets with p$\mathrm{_T\gtrsim 100}$ GeV. 

\begin{figure}[H]
  \subfigure{\includegraphics[width=0.85\linewidth]{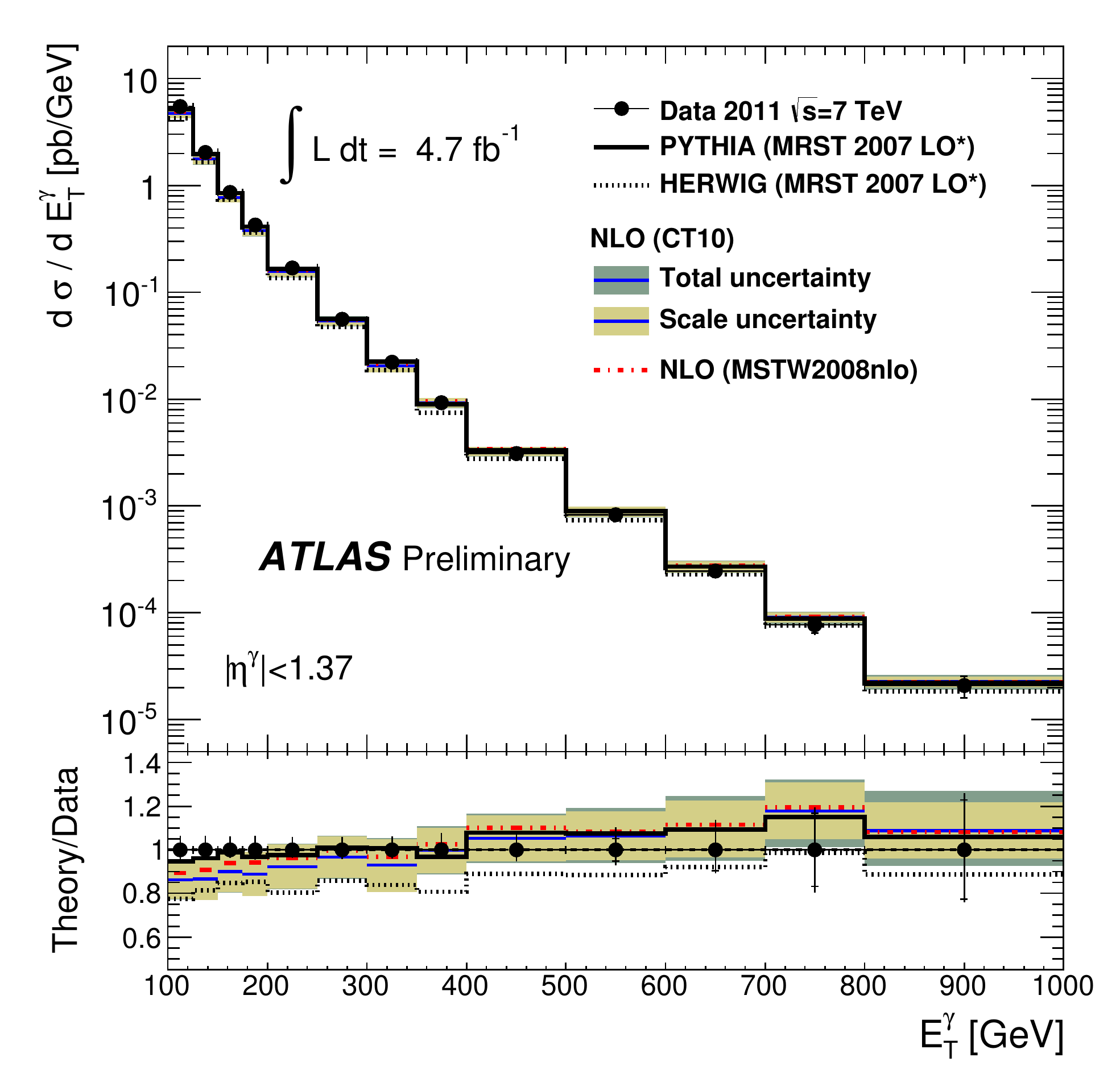}}
  \caption{Measured and expected inclusive prompt photon cross section~\cite{ATLAS-CONF-2013-022} in the barrel $\eta$ region ($|\eta^\gamma | <1.37$). 
    The inner error bars on the data points show statistical uncertainties, while the full error bars show statistical and systematic uncertainties added in quadrature. 
    The NLO QCD prediction is shown as a shaded band which indicates theoretical uncertainties, 
    while the LO parton shower Monte Carlo generators ({\sc{Pythia}}, {\sc{Herwig}}~\cite{Corcella:2002jc}) are shown as lines. \label{fig:IncPhotons}}
\end{figure}
\begin{figure*}[th]
    \subfigure{\includegraphics[width=0.32\linewidth]{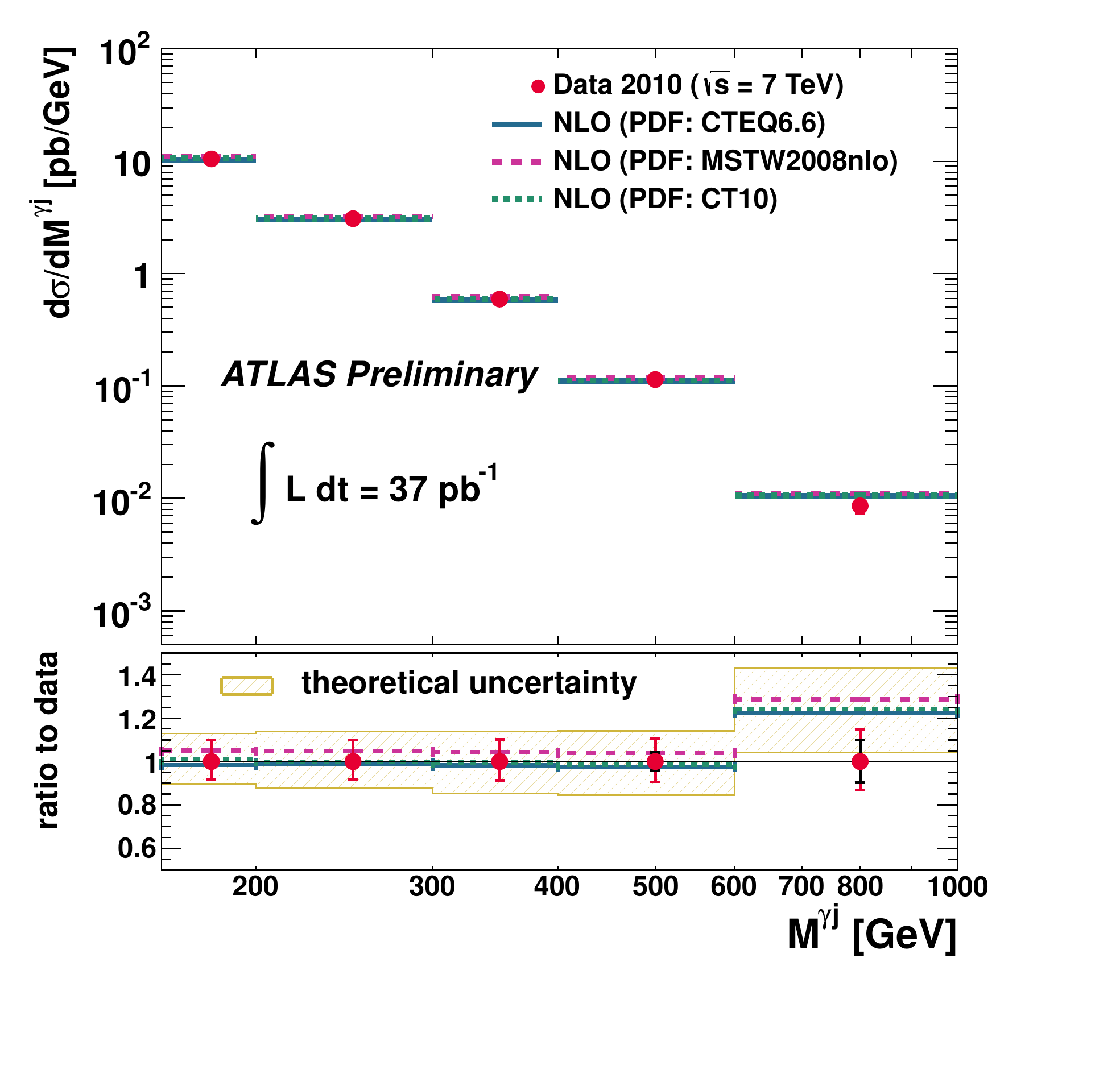}}  
    \subfigure{\includegraphics[width=0.32\linewidth]{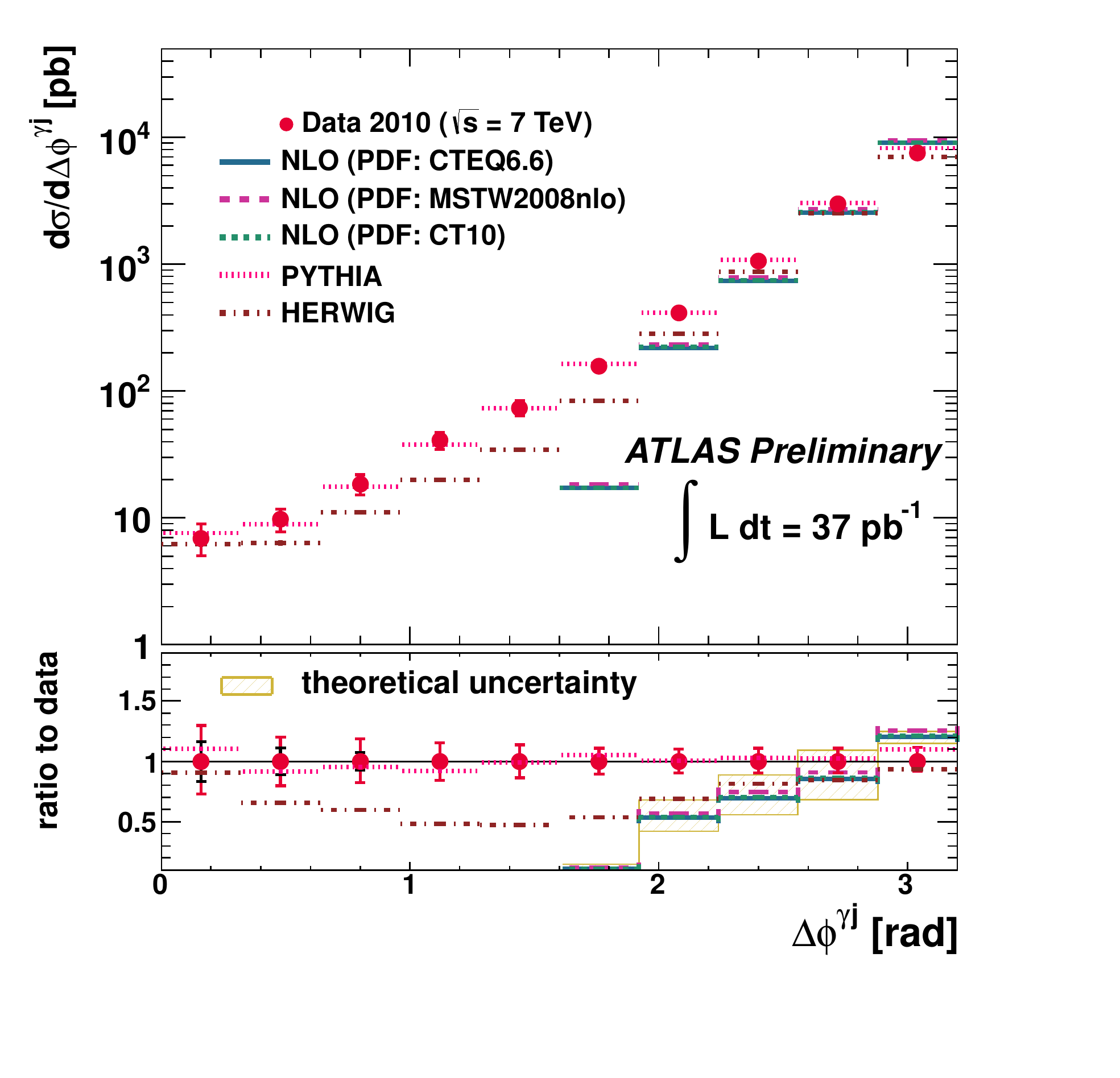}}  
    \subfigure{\includegraphics[width=0.32\linewidth]{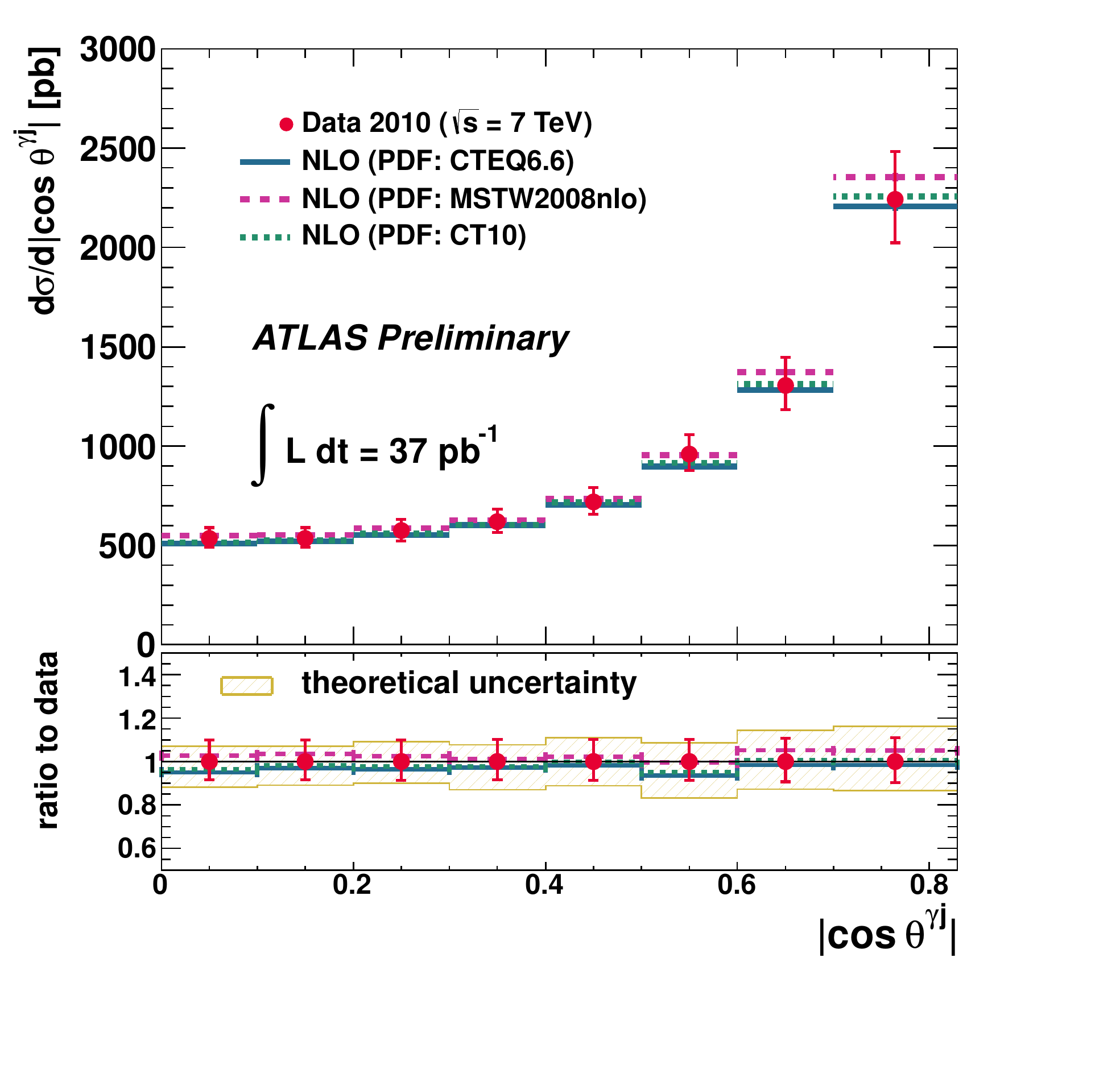}}  
    \caption{The measured differential cross section for isolated-photon plus jet production (dots) as a function of M$^{\mathrm{ \gamma j }}$, $\mathrm{ \Delta\phi^{\gamma j} }$ 
    and $\mathrm{ |\cos\theta^{\gamma j}| }$~\cite{ATLAS-CONF-2013-023}. 
    The NLO QCD calculations from {\sc{Jetphox}} using the 
    {\sc{Cteq6.6}} (solid lines), {\sc{Mstw2008}} (dashed lines) and {\sc{Ct10}} (dotted lines) PDF sets are also shown. 
    The bottom part of the figure shows the ratio between the NLO QCD calculations and the measured cross section. 
    The inner (outer) error bars represent the statistical uncertainties (the statistical and systematic uncertainties added in quadrature) 
    and the hatched band represents the theoretical uncertainty.\label{fig:PhotPlusJets}}
\end{figure*}

\section{Isolated photon measurement\label{sec:Photons}}

Measurement of the isolated prompt photon cross section provides a test of the QCD predictions with the good experimental precision that can be achieved using electroweak probes 
such as photons. The cross section definition at particle level includes photons produced by fragmentation but exclude meson decay products, 
such as $\mathrm{\pi^0\rightarrow \gamma\gamma}$ and $\mathrm{ \eta^0\rightarrow \gamma\gamma}$, which are treated as background.

The inclusive cross section for photon production was recently reported in Ref.~\cite{ATLAS-CONF-2013-022}; 
in Fig.~\ref{fig:IncPhotons} the transverse energy spectrum $\mathrm{ E_T^\gamma }$ of the photons 
is shown along with LO Monte Carlo generators, {\sc{Pythia}} and {\sc{Herwig}}, and NLO QCD predictions, {\sc{Jetphox}}~\cite{Bern:2002jx}. 

More complex distributions involving photon production in association with high-$\mathrm{ p_T }$ jets have been measured~\cite{ATLAS-CONF-2013-023}.  
In Fig.~\ref{fig:PhotPlusJets} we show few selected results, namely the invariant mass (M$\mathrm{ ^{\gamma j} }$), the azimuthal angle separation 
($\mathrm{\Delta\phi^{\gamma j}}$) and the 
absolute value of the cosine of the Collins-Soper angle ($\mathrm{|\cos\theta^{\gamma j}|}$~\cite{Collins:1977iv}) of the photon-jet system. 
The data are well described by NLO QCD predictions obtained with {\sc{Jetphox}}, 
but LO parton shower MC provide a more accurate description of the $\mathrm{\Delta\phi^{\gamma j}}$ distribution for $\mathrm{\Delta\phi^{\gamma j}\lesssim 2}$.

\begin{figure*}[th]
  \subfigure{\includegraphics[width=0.32\linewidth]{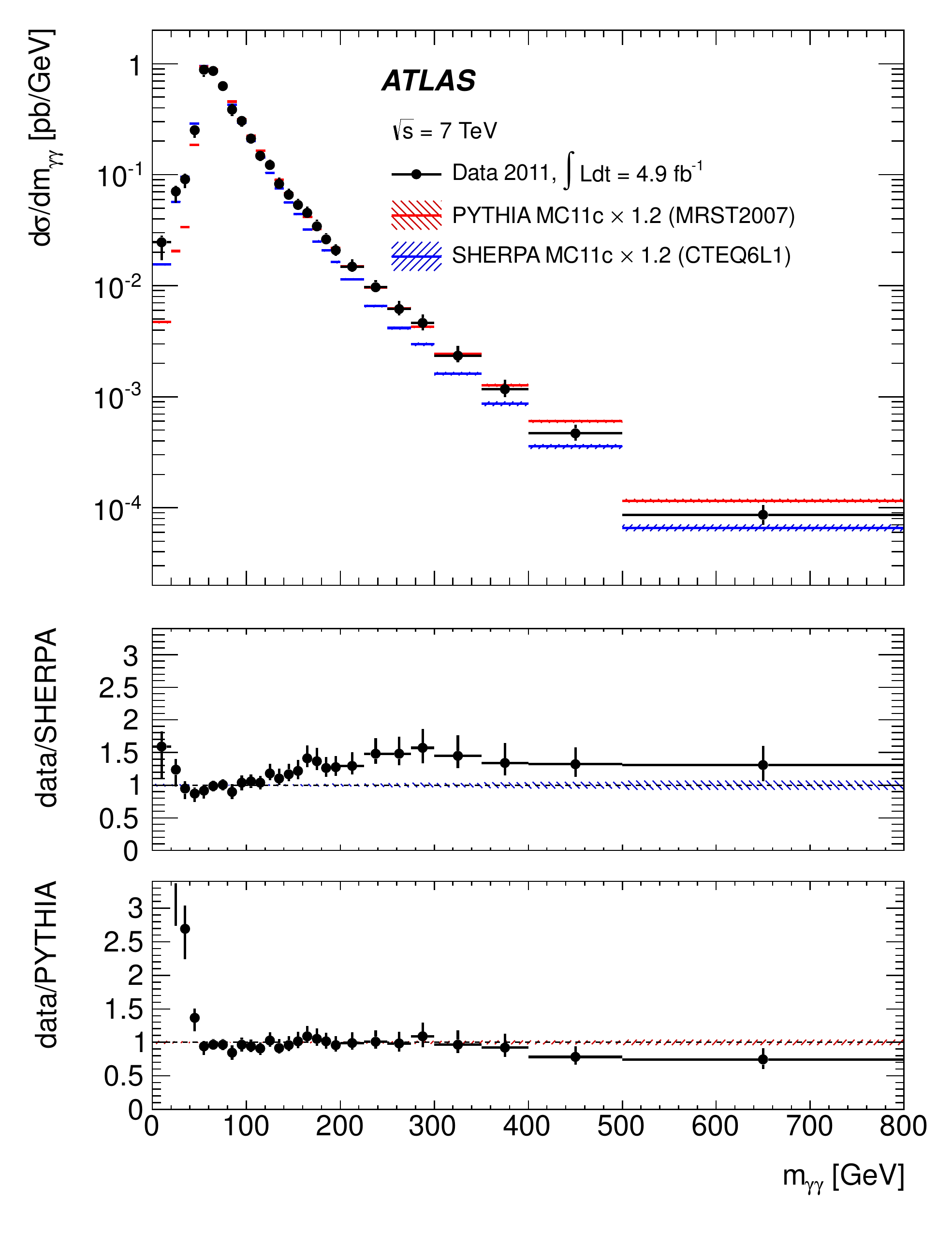}}
  \subfigure{\includegraphics[width=0.32\linewidth]{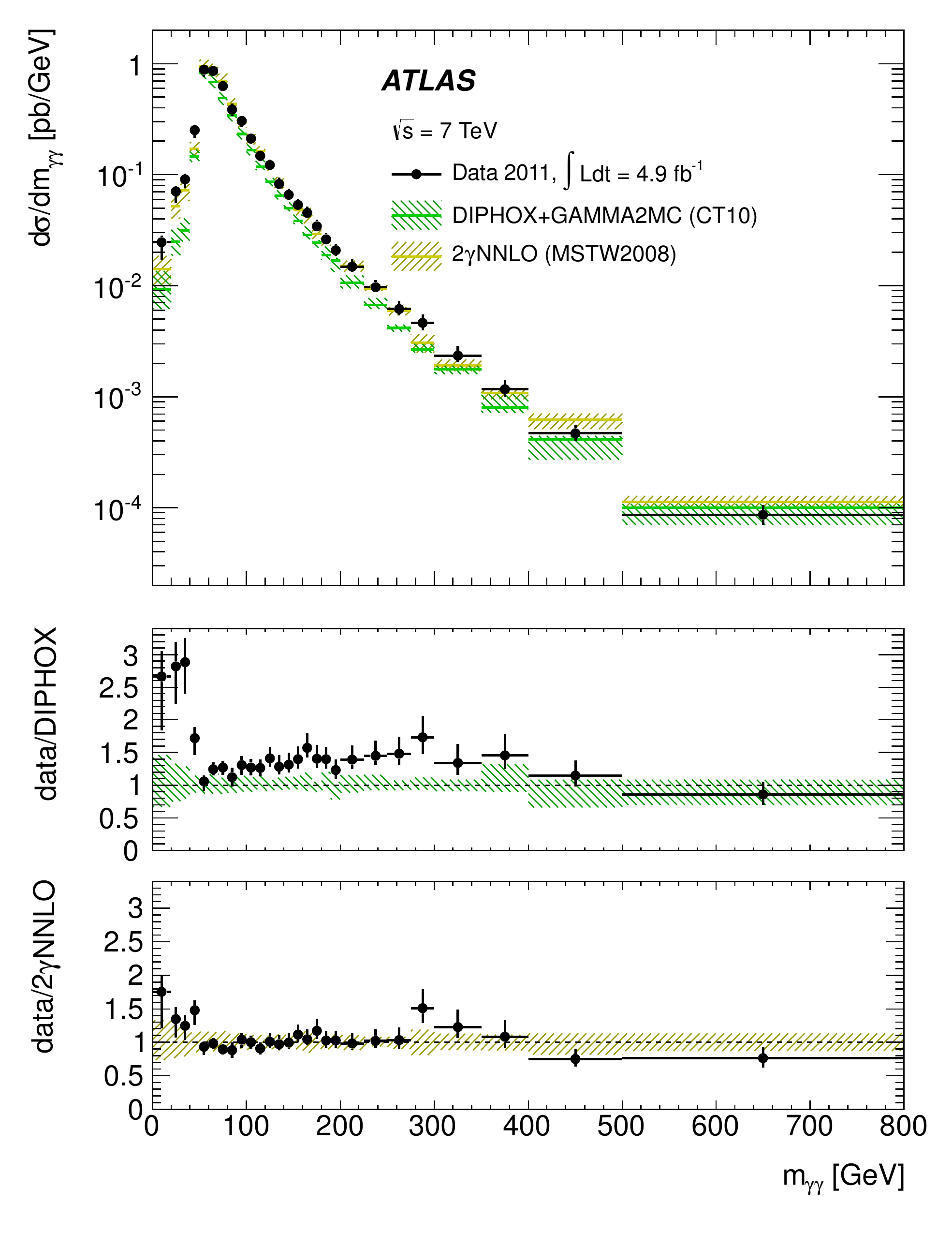}}
  \subfigure{\includegraphics[width=0.32\linewidth]{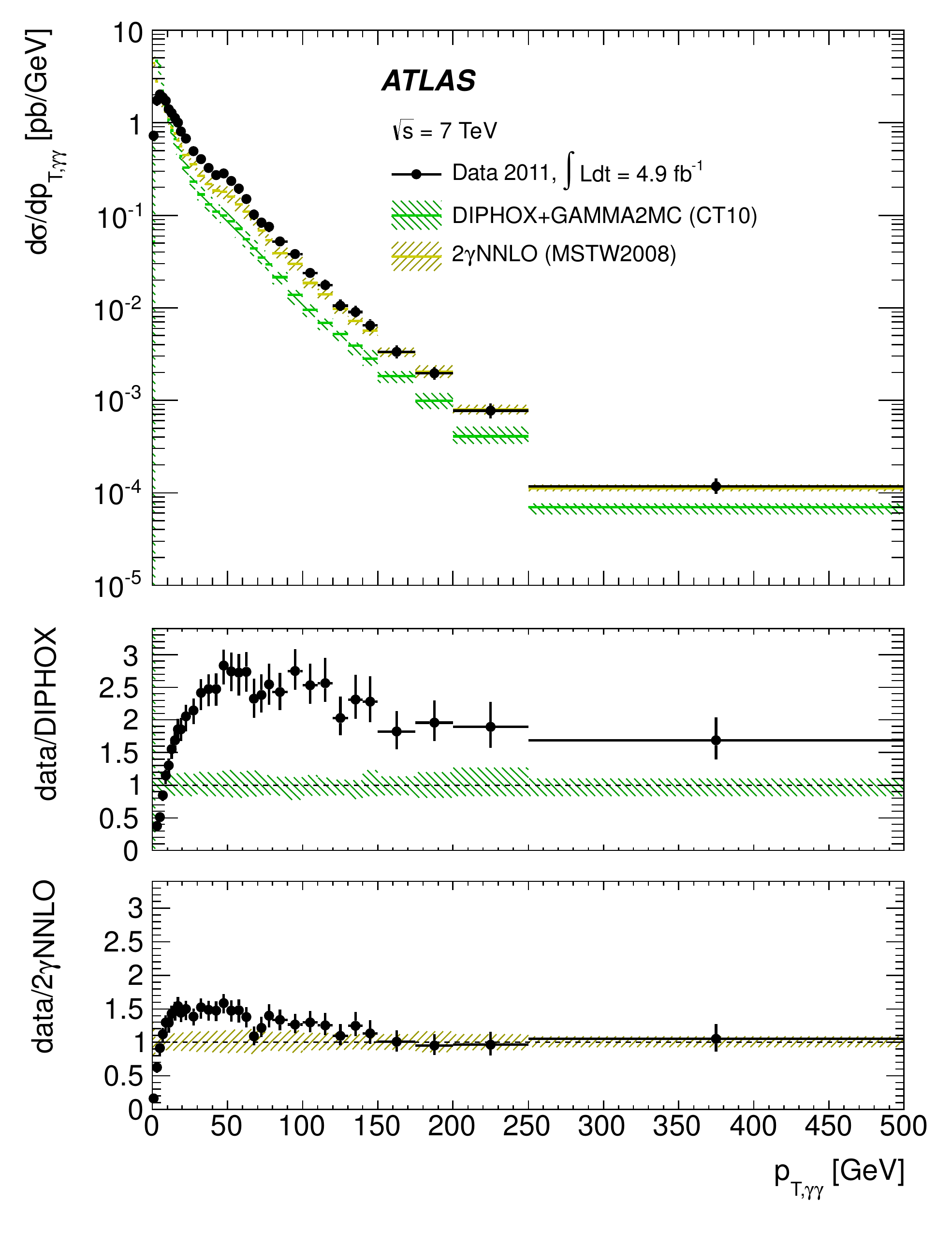}}                   
  \caption{Invariant mass distribution of the leading photon pairs m$\mathrm{_{\gamma\gamma}}$ (left and center) and 
    transverse momentum p$\mathrm{_{T,\gamma\gamma}}$ of the di-photon system (right)~\cite{Aad:2012tba}.  
    Black dots correspond to data with error bars for their total uncertainties.
    The data are compared to LO MC generators (left), {\sc{Pythia}} and {\sc{Sherpa}}, NNLO, {\sc{2$\mathrm{\gamma}$nnlo}}, and NLO, {\sc{Diphox++Gamma2MC}} calculations 
    (center and right). The uncertainty on the cross sections predicted by LO MC generators include only statistical uncertainties, 
    while uncertainties on the NLO and NNLO QCD predictions include contributions from the limited size of the simulated sample, 
    from the scale choice and from uncertainties on the parton distribution functions and on the hadronization and underlying event corrections. \label{fig:DiPhotons}}
\end{figure*}

Of particular interest are the measurements of the di-photon system which accounts for the largest and irreducible background 
in Higgs boson selection in the decay channel 
$\mathrm{H\rightarrow \gamma\gamma}$ and in BSM di-photon resonance searches. 
Using the 2011 dataset, the differential distribution of the invariant mass ($\mathrm{m_{\gamma\gamma}}$) and transverse momentum ($\mathrm{p_{T,\gamma\gamma}}$) 
have been measured in~\cite{Aad:2012tba}.
The data are compared to LO MC generators, {\sc{Pythia}} and {\sc{Sherpa}}, as well as NNLO, {\sc{2$\mathrm{\gamma}$nnlo}}, and NLO QCD calculations, 
{\sc{Diphox+Gamma2MC}}~\cite{Binoth:1999qq,Bern:2002jx}.
Fig.~\ref{fig:DiPhotons} shows the invariant mass distribution (left and center) and transverse momentum of the di-photon system (right).  
The total cross sections of LO MC generators have been normalized to the data cross section, while the NNLO and NLO QCD predictions are compared with data 
without cross section rescaling. The best description of the data is provided by NNLO QCD predictions.


\section{Inclusive massive gauge boson production\label{sec:WZ}}

\begin{figure*}[th]
  \begin{tabular}{ccc}
    \subfigure{\includegraphics[width=0.32\linewidth]{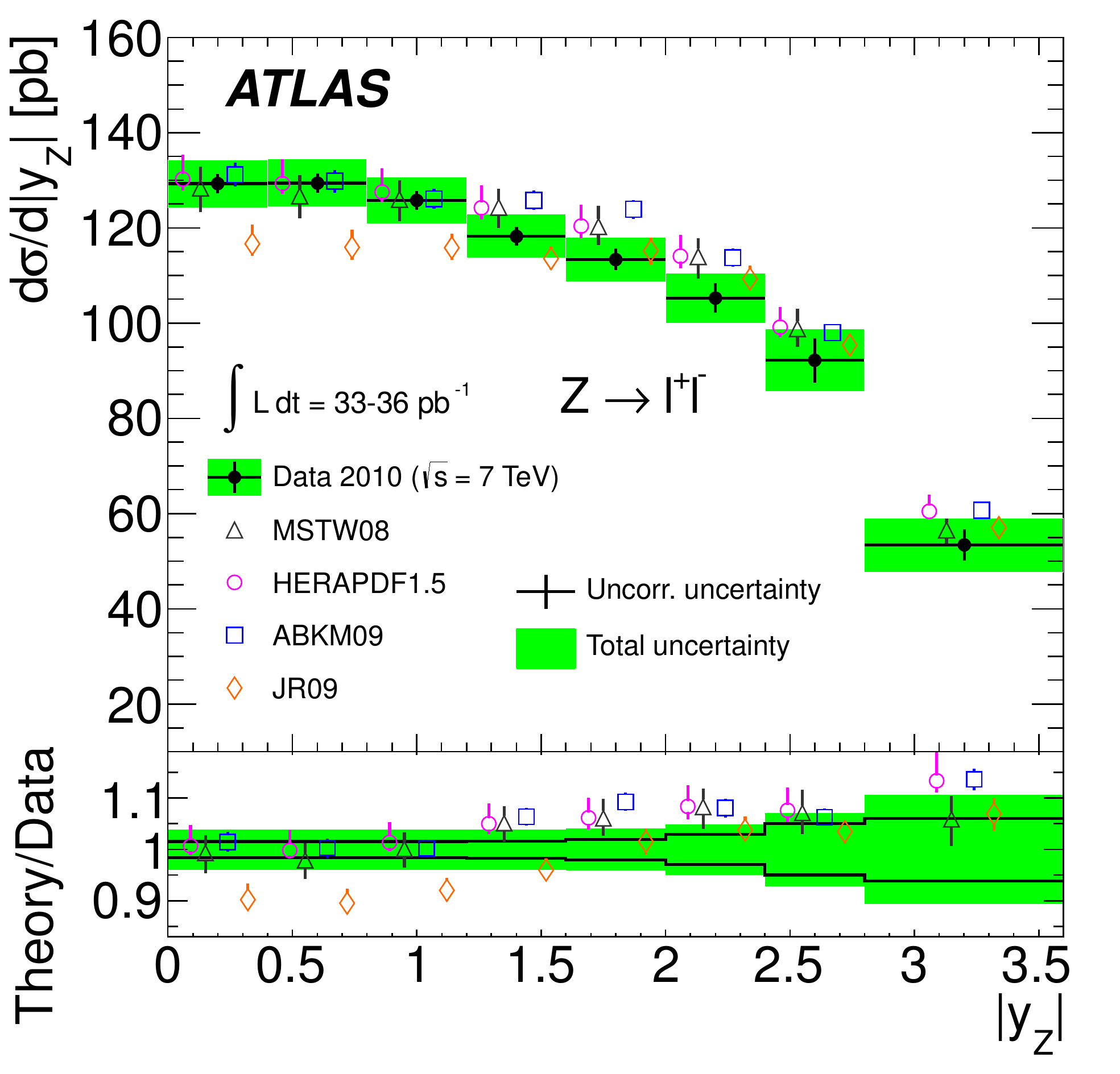}}  &
    \subfigure{\includegraphics[width=0.32\linewidth]{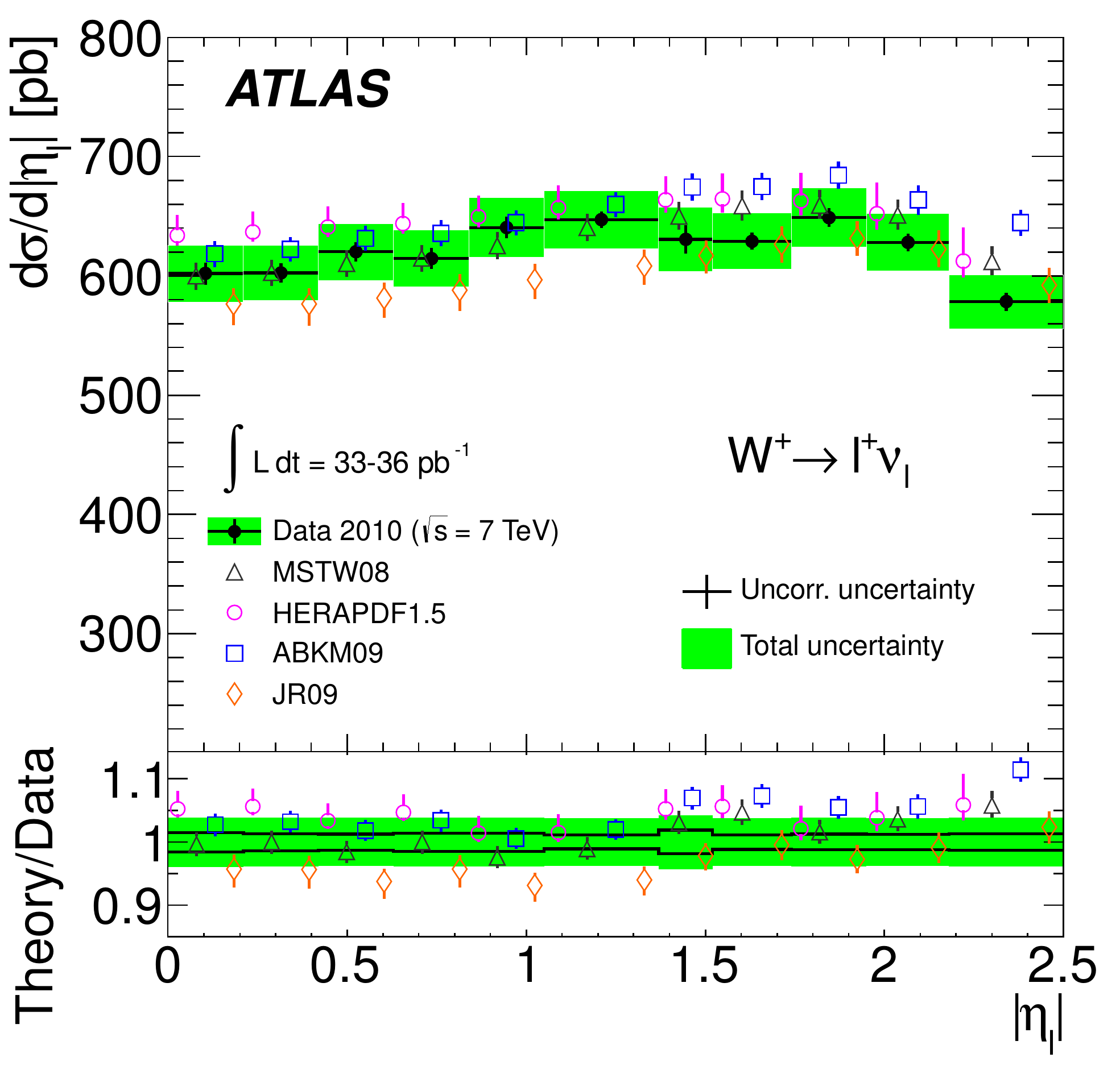}} &
    \subfigure{\includegraphics[width=0.32\linewidth]{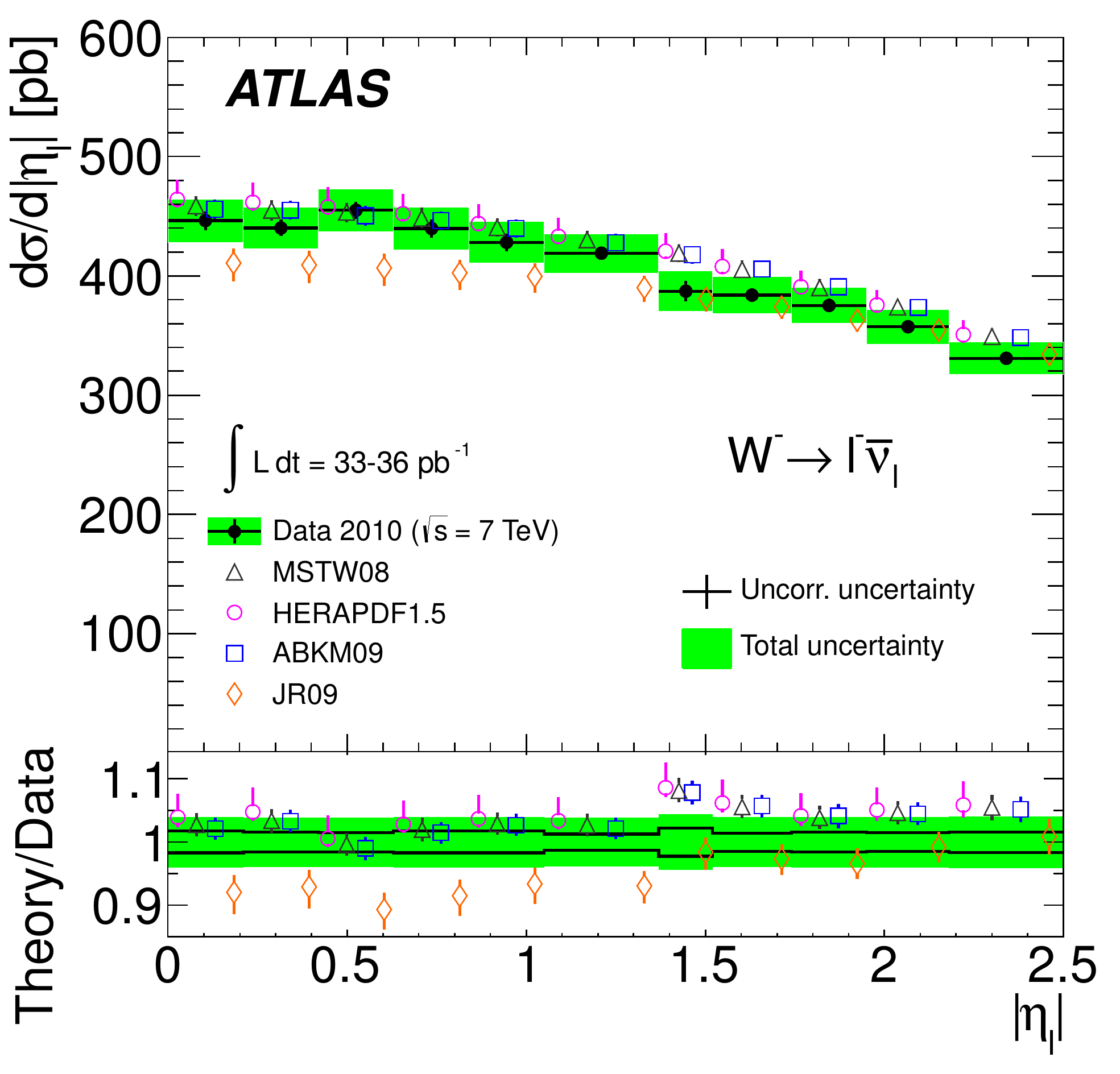}} \\
  \end{tabular}
  \caption{Differential production cross section measurement for $\mathrm{Z\rightarrow \ell\ell}$ as a function of $\mathrm{|y_Z|}$ (left) and for 
    $\mathrm{W^{\pm}\rightarrow \ell^{\pm}\nu}$ as a function of $\mathrm{|\eta_{\ell^{\pm}}|}$ (center and right)~\cite{Aad:2011dm}. 
    The data are compared to NNLO QCD predictions ({\sc{Fewz}} and {\sc{Dynnlo}}) using various PDF sets. 
    \label{fig:InclusiveWZ}}
\end{figure*}
\begin{figure}[th]
    \subfigure{\includegraphics[width=0.8\linewidth]{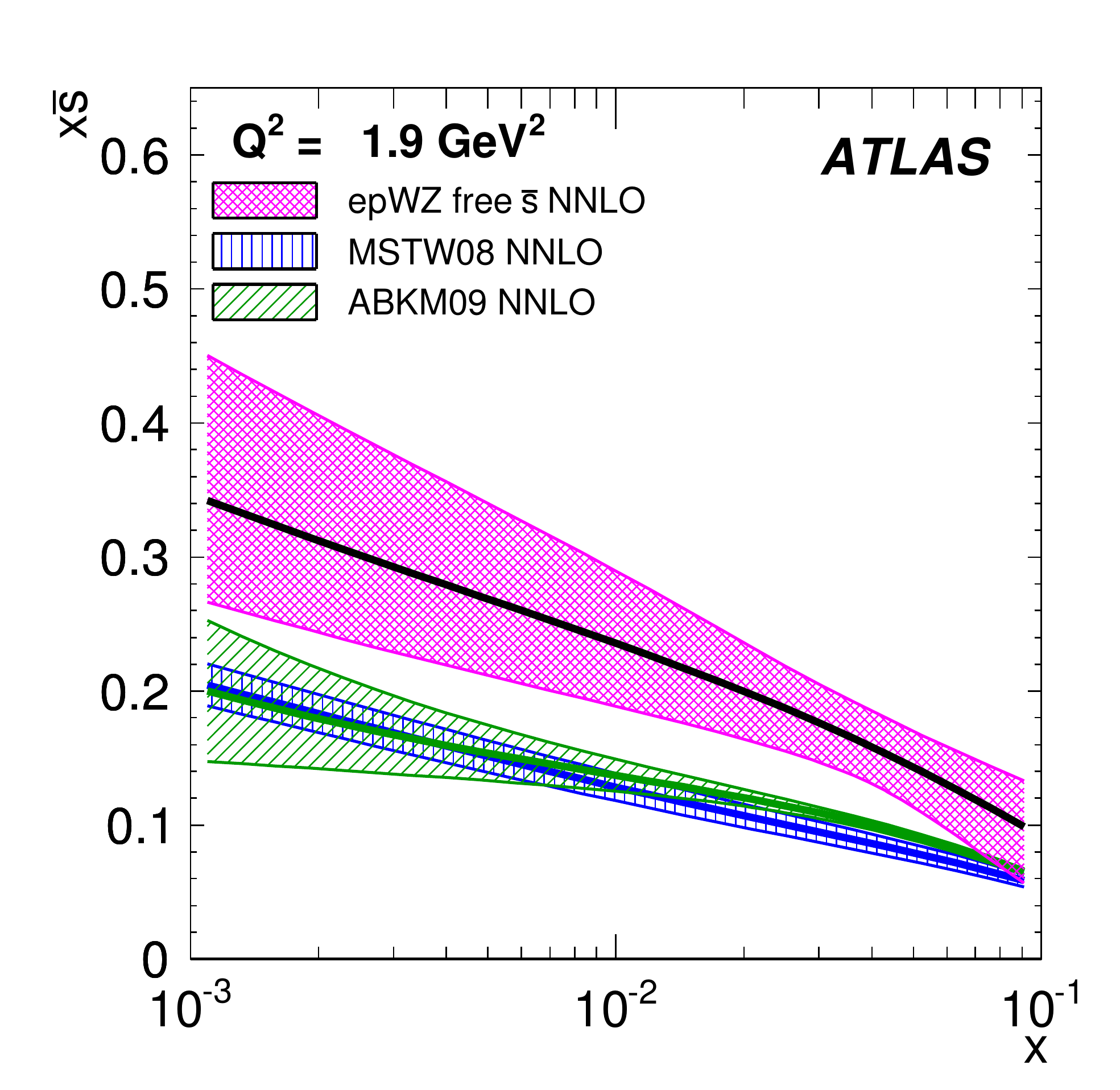}}
    \caption{The strange anti-quark density as a function of x for the epWZ free $\mathrm{\overline s}$ NNLO fit (magenta band) compared to predictions from {\sc{Mstw2008}} 
      (blue hatched) and {\sc{Abkm09}} (green hatched) at $\mathrm{Q^2= 1.9}$ GeV$^2$~\cite{Aad:2012sb}.\label{fig:WZPDF}}
\end{figure}

\begin{figure}[th]
  \subfigure{\includegraphics[width=0.96\linewidth]{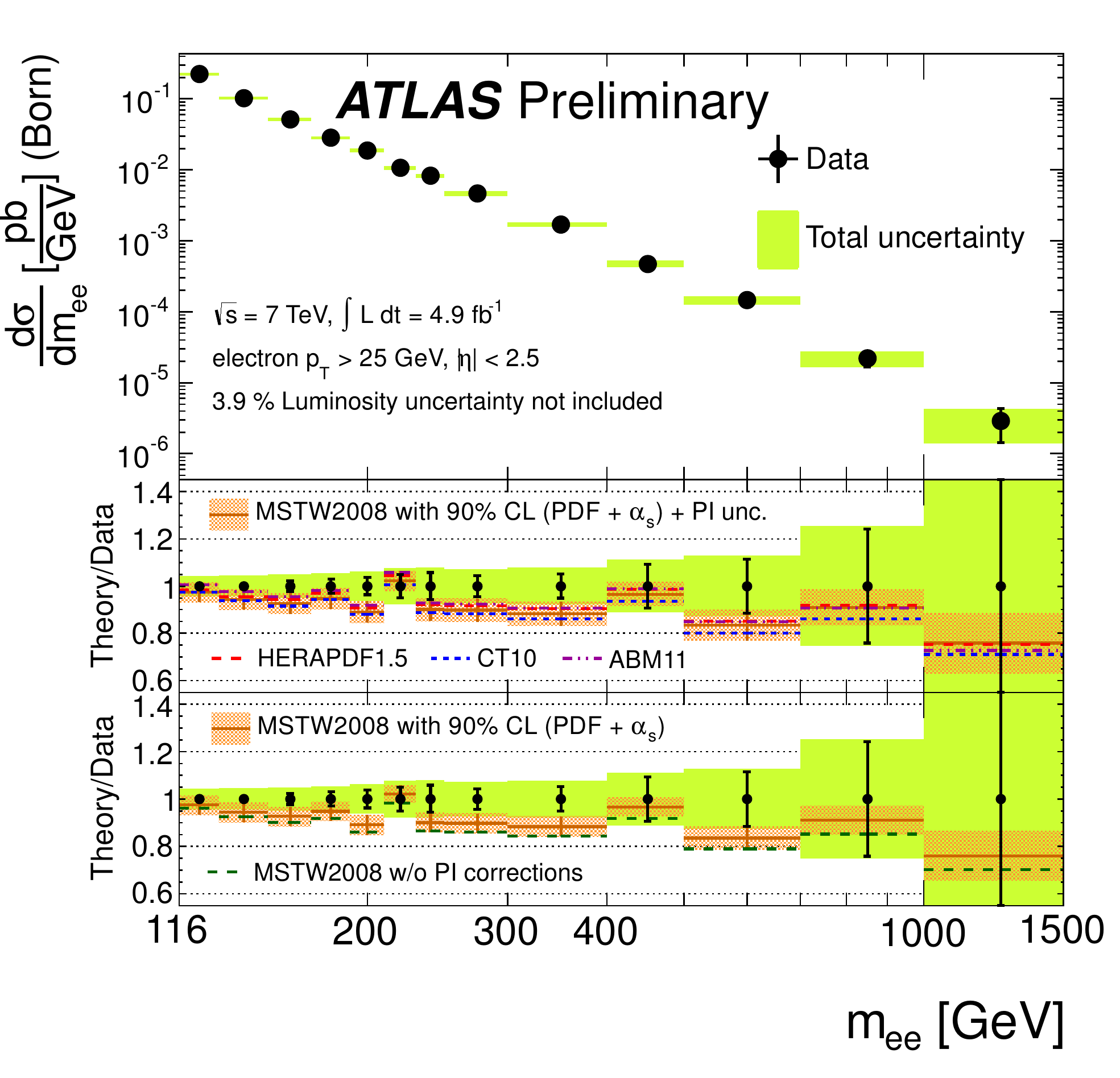}}
  \caption{Measured differential cross-section~\cite{ATLAS-CONF-2012-159} 
    with statistical and combined statistical and systematic uncertainties, excluding the $3.9\%$ uncertainty on the luminosity. 
    The measurement is compared to {\sc{Fewz3.1}} calculations at NNLO using the $\mathrm{ G_\mu }$ electroweak parameter scheme and including electroweak corrections. 
    In the upper ratio plot, the photon-induced (PI) corrections have been added to the predictions of the {\sc{Mstw2008}}, 
    {\sc{Ct10}}, {\sc{Herapdf1.5}} and {\sc{Abm11}} NNLO PDFs, and for the {\sc{Mstw2008}} prediction, the total uncertainty band arising from the PDF, $\mathrm{ \alpha_s }$ 
    and photon-induced uncertainties is drawn. 
    The lower ratio plot shows the influence of the photon-induced corrections on the {\sc{Mstw2008}} prediction, 
    the uncertainty band including only the PDF and $\mathrm{ \alpha_s }$ uncertainties.\label{fig:HighMassDY}}
\end{figure}
\begin{figure}[th]
  \subfigure{\includegraphics[width=0.84\linewidth]{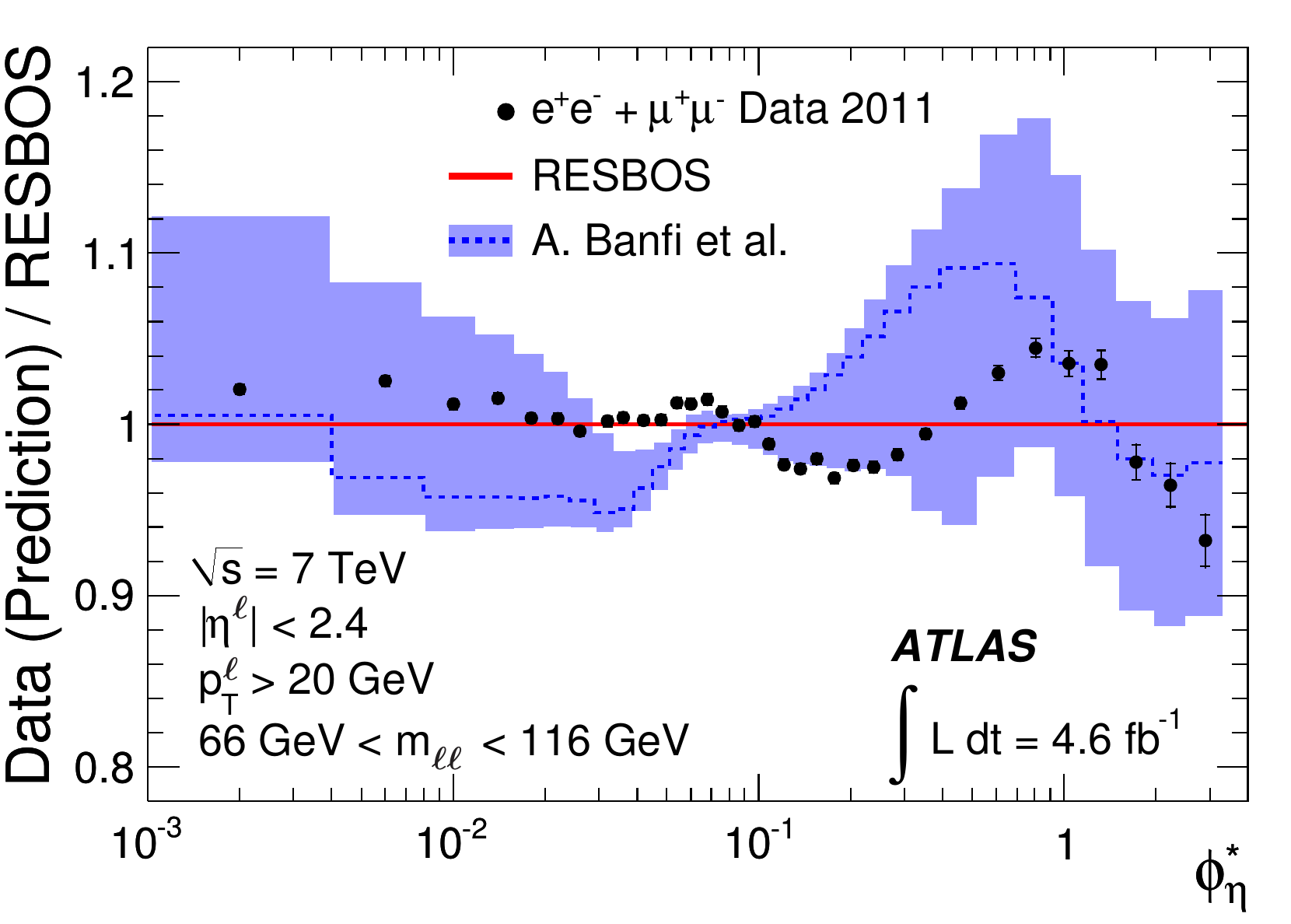}}
  \subfigure{\includegraphics[width=0.84\linewidth]{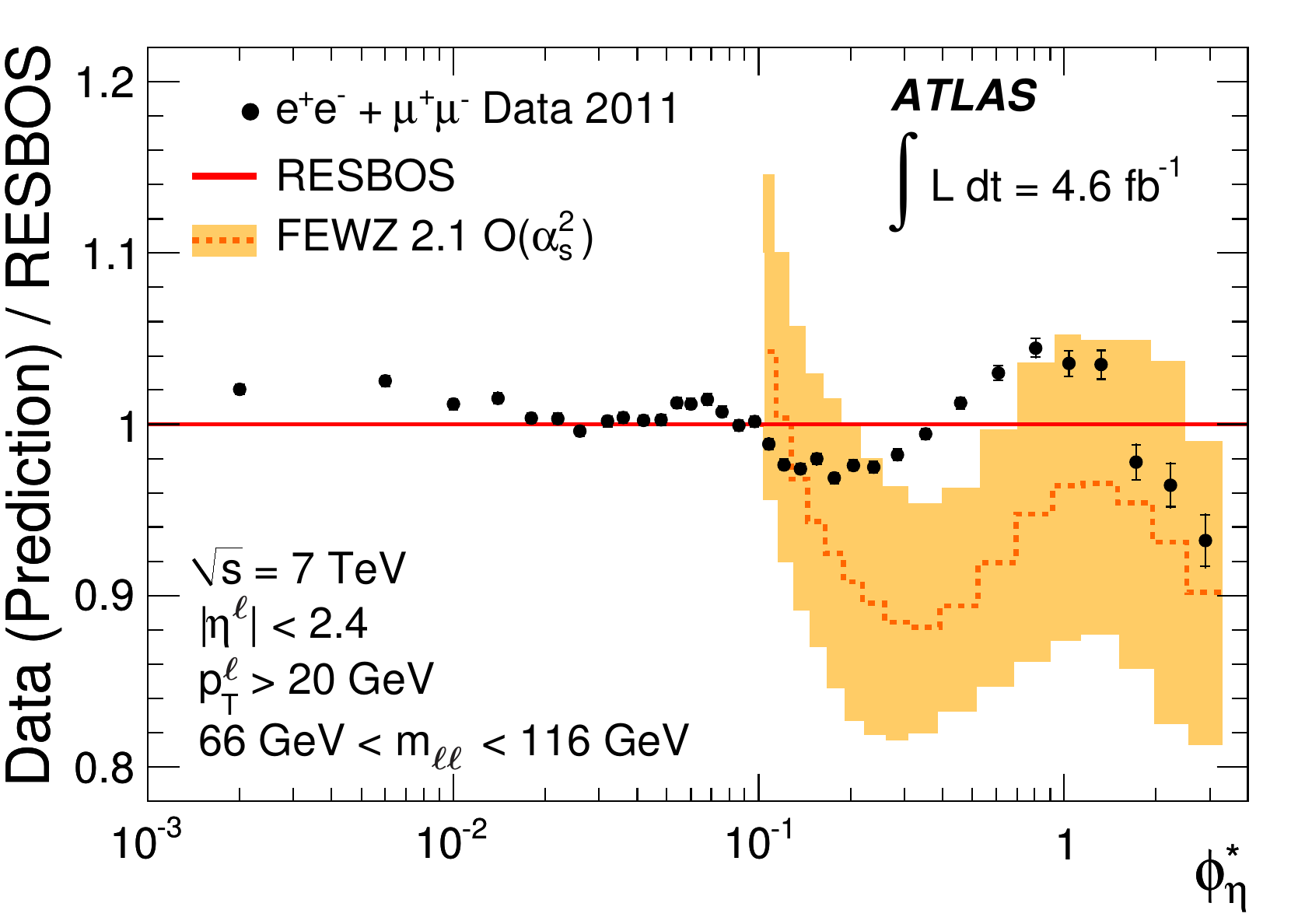}}
  \caption{The ratio of the measured normalized differential cross section~\cite{Aad:2012wfa} to {\sc{ResBos}} predictions as a function of $\mathrm{ \phi_\eta^* }$. 
    The inner and outer error bars on the data points represent the statistical and total uncertainties, respectively. 
    The measurement is also compared to predictions, which are represented by a dashed line, 
    from A.~Banfi, et al. Phys.\ Lett.\ B {\bf 715} (2012) 152 (top) and to NNLO QCD predicition obtained with {\sc{Fewz2.1}} (bottom). 
    Uncertainties associated to this calculation are represented by a shaded band. The prediction from {\sc{Fewz2.1}} is only presented for 
    $\mathrm{ \phi_\eta^*>0.1 }$.\label{fig:PhiStar}}
\end{figure}

\begin{figure*}[ht]
  \subfigure{\includegraphics[width=0.32\linewidth]{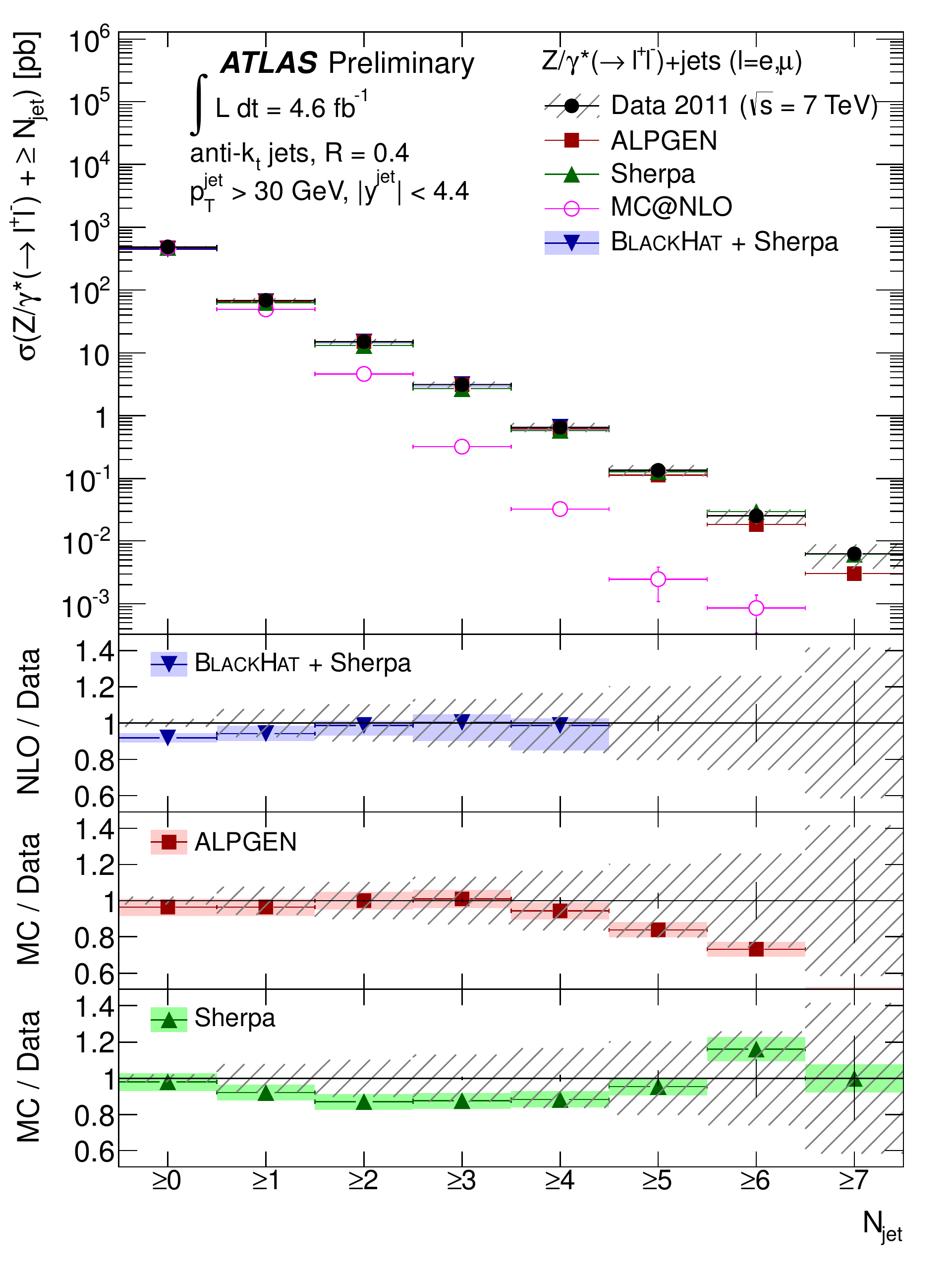}}
  \subfigure{\includegraphics[width=0.32\linewidth]{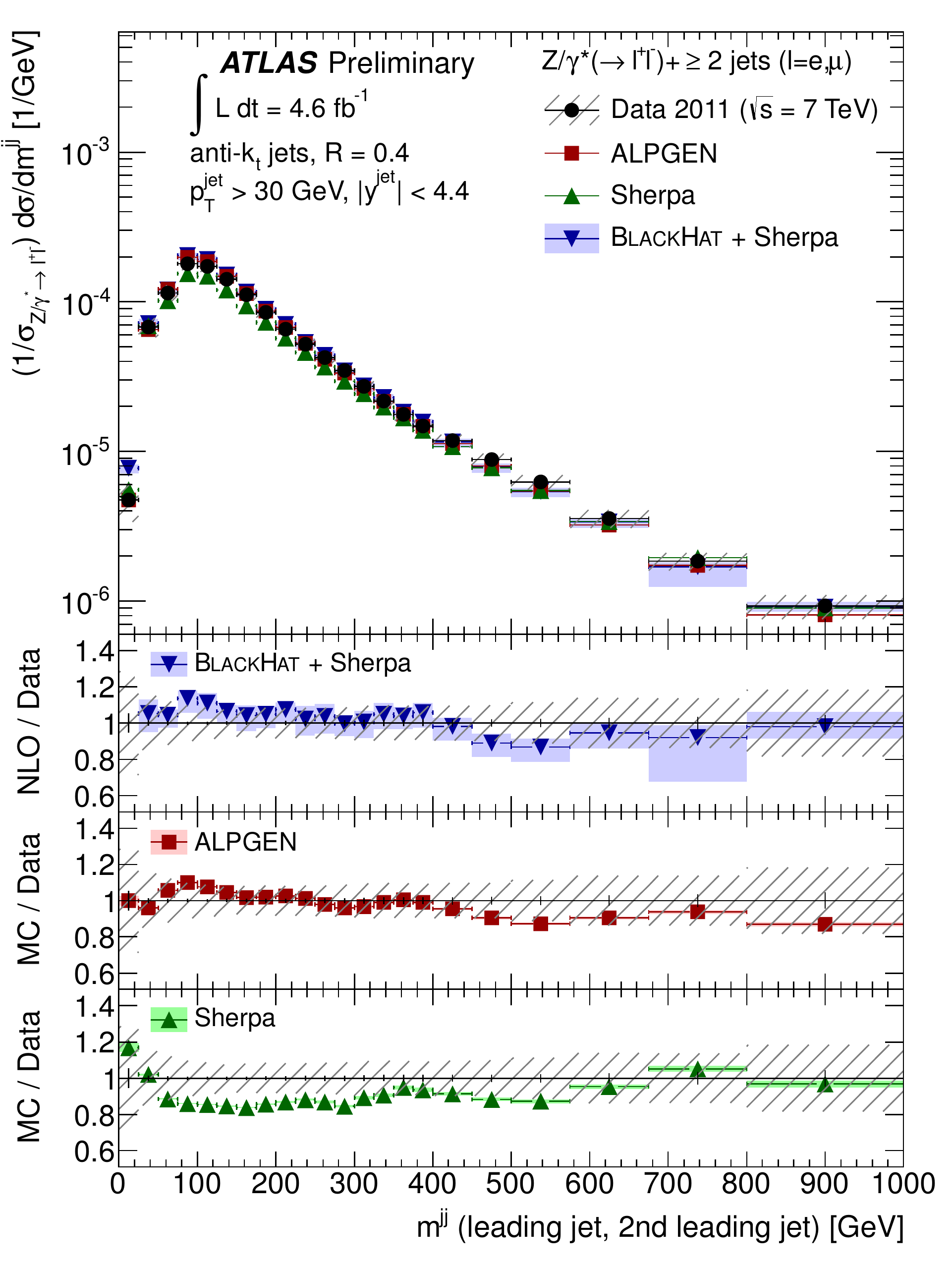}}
  \subfigure{\includegraphics[width=0.32\linewidth]{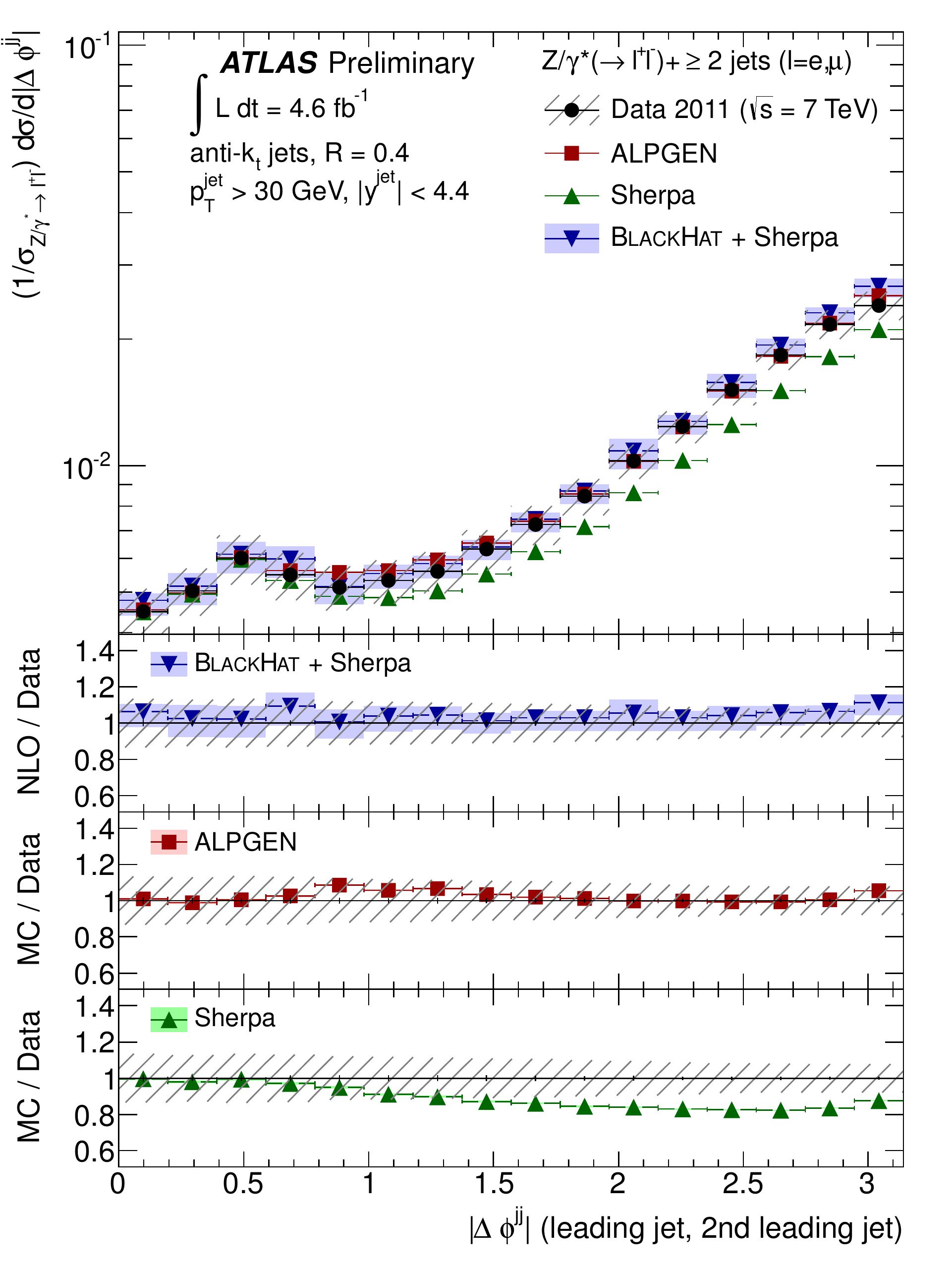}}
  \caption{Measured cross section for Z($\rightarrow\ell\ell$)+jets as a function of the inclusive jet multiplicity (left), invariant mass of the two leading jets (center) and 
    $\mathrm{\Delta\phi}$ between the two leading jets (right)~\cite{Z+jets}.
    The data are compared to NLO multileg QCD predictions {\sc{Blackhat+Sherpa}} and the {\sc{Alpgen}}, {\sc{Sherpa}} and {\sc{Mc@nlo}} event generators. 
    The error bars indicate the statistical uncertainty on the data, and the hatched (shaded) bands the statistical and systematic uncertainties on data (theory). 
    \label{fig:Z+jets}}
\end{figure*}
\begin{figure}[tbp]
  \subfigure{\includegraphics[width=0.85\linewidth]{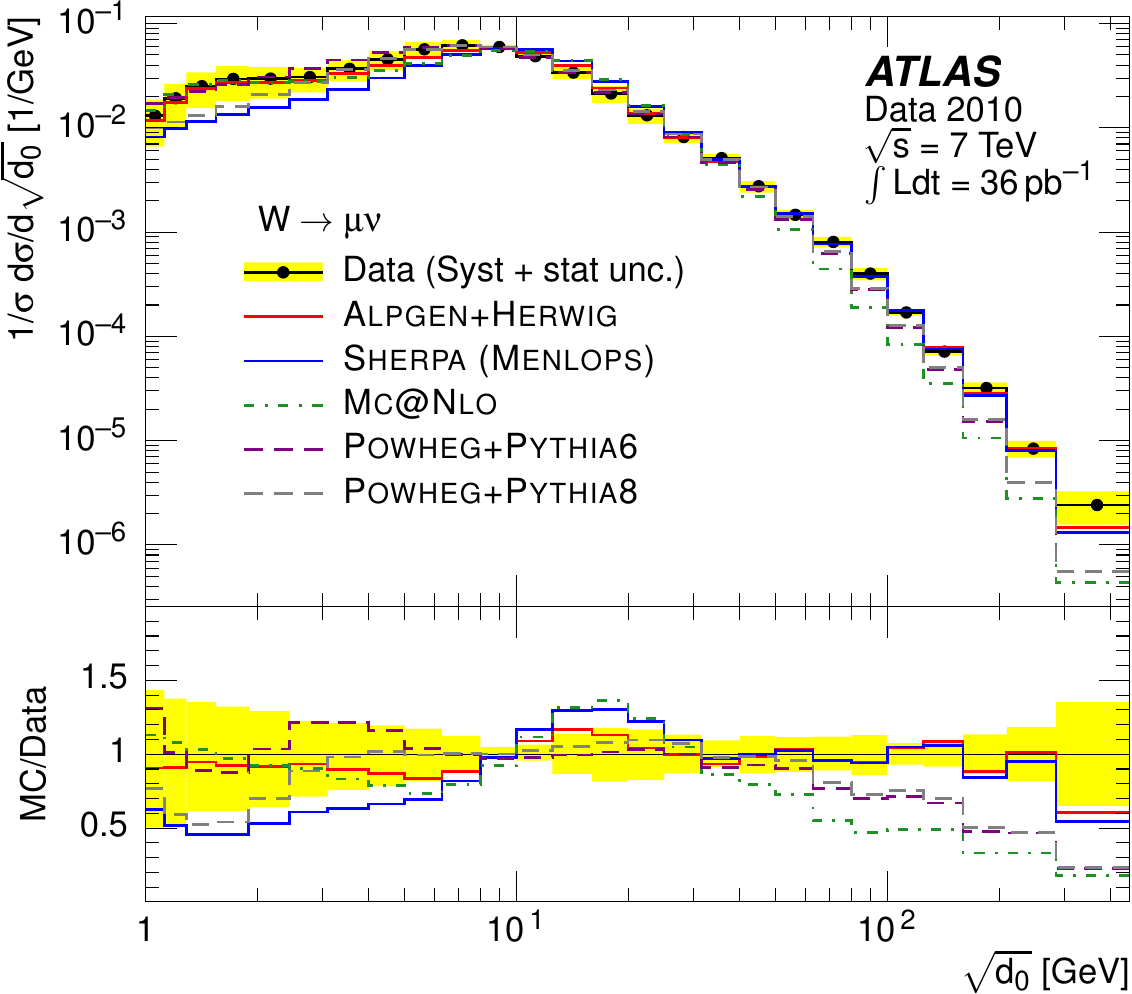}}  
  \caption{Distribution of $\mathrm{\sqrt{d_0}}$ in the $\mathrm{W\rightarrow \mu\nu}$ channel~\cite{Aad:2013ueu}.
    The data (markers) are compared to the predictions from various MC generators and the shaded bands represent the quadrature sum of systematic and 
    statistical uncertainties on each bin. The distribution has been normalized to unity.\label{fig:SplittingScales}}
\end{figure}
\begin{figure}[ht]
  \subfigure{\includegraphics[width=0.85\linewidth]{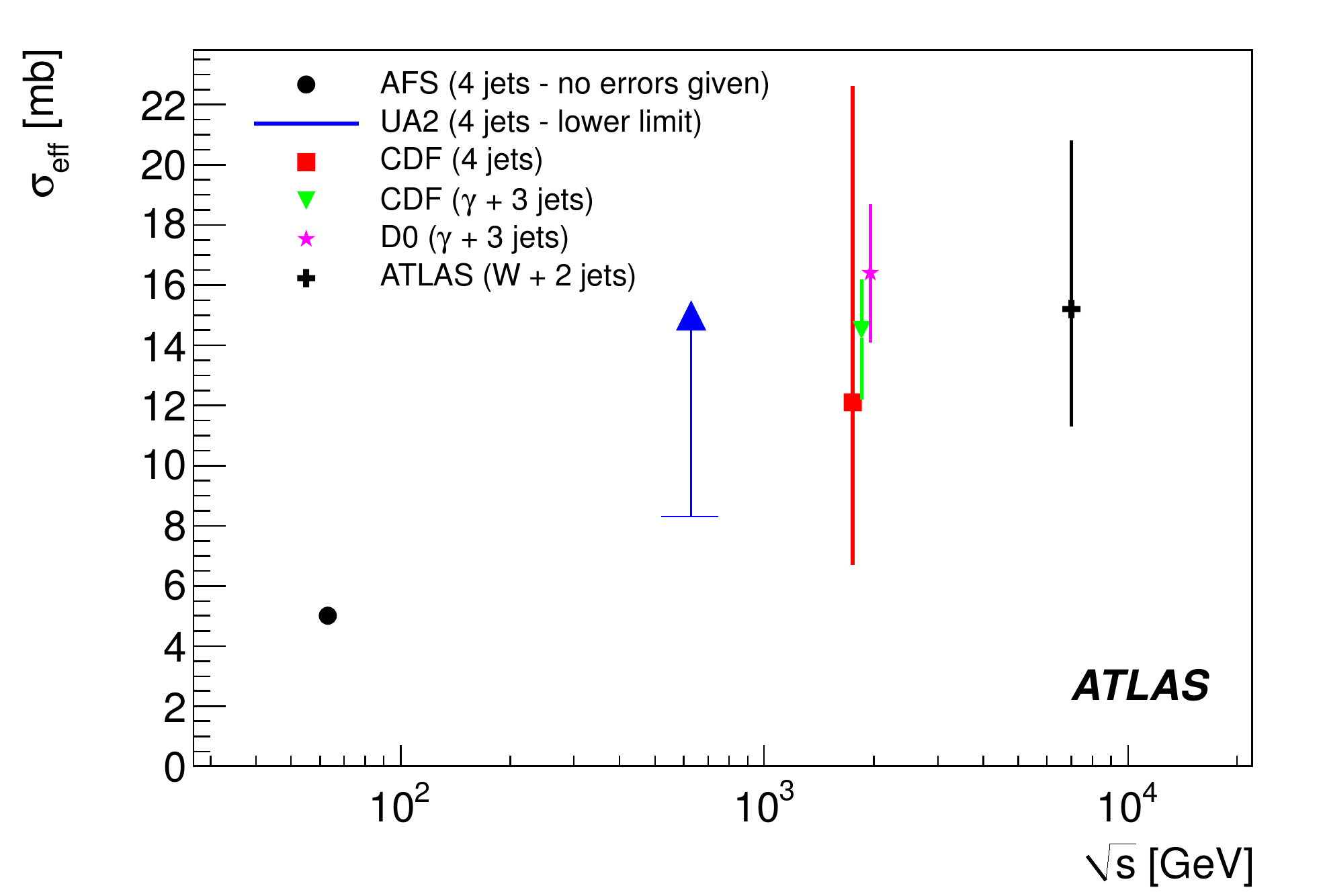}}
  \caption{The center-of-mass energy $\mathrm{ \sqrt s }$ dependence of $\mathrm{ \sigma_{\mathit{eff}} }$~\cite{Aad:2013bjm} extracted in different processes and experiments. 
    An offset has been applied to the 1.8TeV data points in order to distinguish them. 
    The error bars on the data points represent the statistical and systematic uncertainties added in quadrature.\label{fig:DPI}}
\end{figure}
\begin{figure}[ht]
  \subfigure{\includegraphics[width=.95\linewidth]{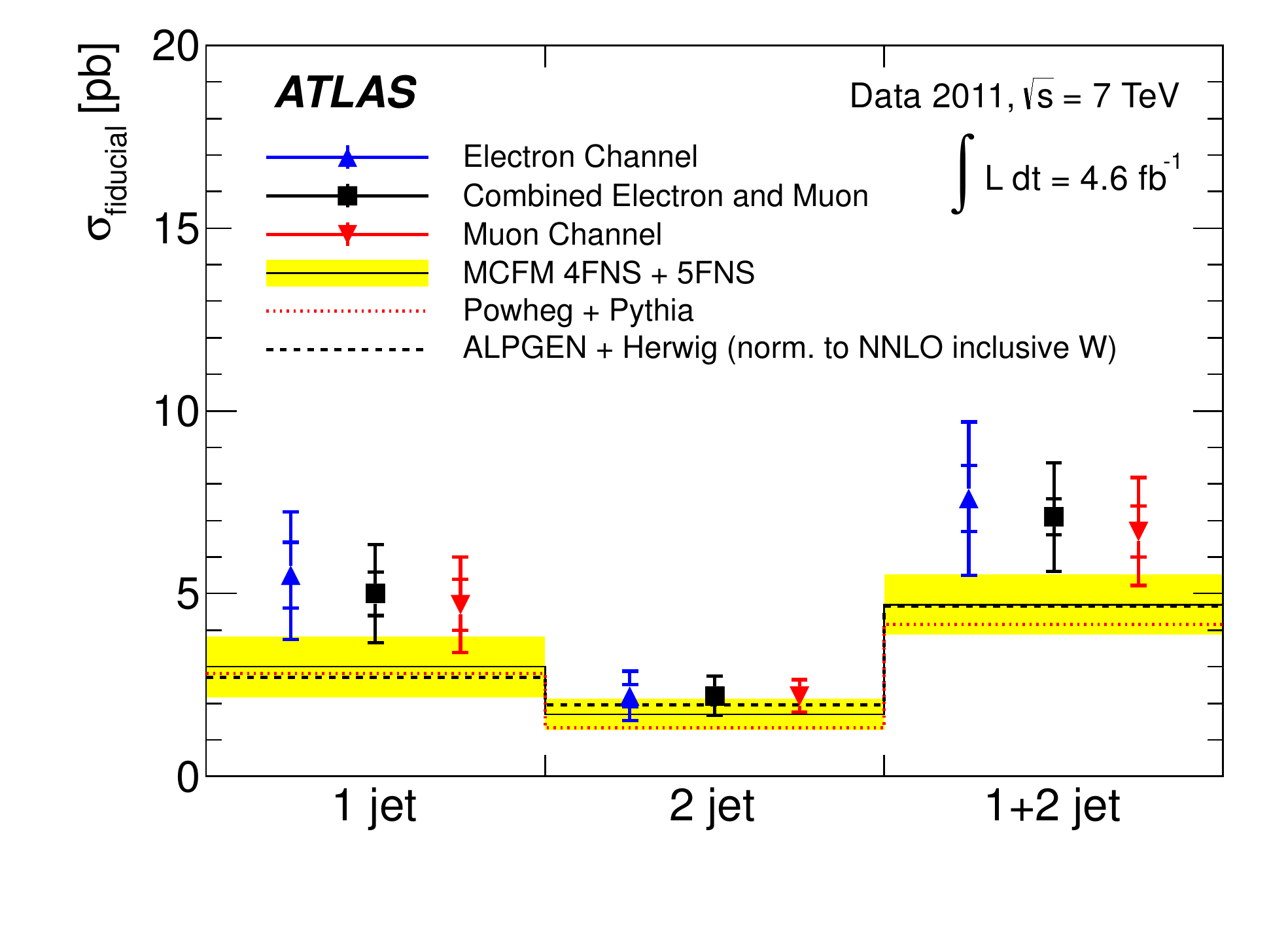}}
  \caption{Measured fiducial cross-sections with the statistical (inner error bar) and statistical plus systematic (outer error bar) uncertainties in the electron, 
    muon, and combined electron and muon channels~\cite{Aad:2013vka}. The cross-sections are given in the 1-jet, 2-jet and 1+2-jet fiducial regions. 
    The measurements are compared with NLO QCD predictions calculated with {\sc{Mcfm}}; the yellow bands represent the total uncertainty on the prediction, obtained 
    by combining in quadrature the uncertainties resulting from variations of the renormalization and factorization scales, 
    the PDF set and the non-perturbative corrections. 
    The NLO prediction from {\sc{Powheg}} interfaced to {\sc{Pythia}} 
    and the prediction from {\sc{Alpgen}} interfaced to {\sc{Herwig}} and {\sc{Jimmy}} (scaled by the NNLO inclusive W normalization factor) are also shown.\label{fig:Wb1}}
\end{figure}
\begin{figure}[ht]
  \subfigure{\includegraphics[width=0.85\linewidth]{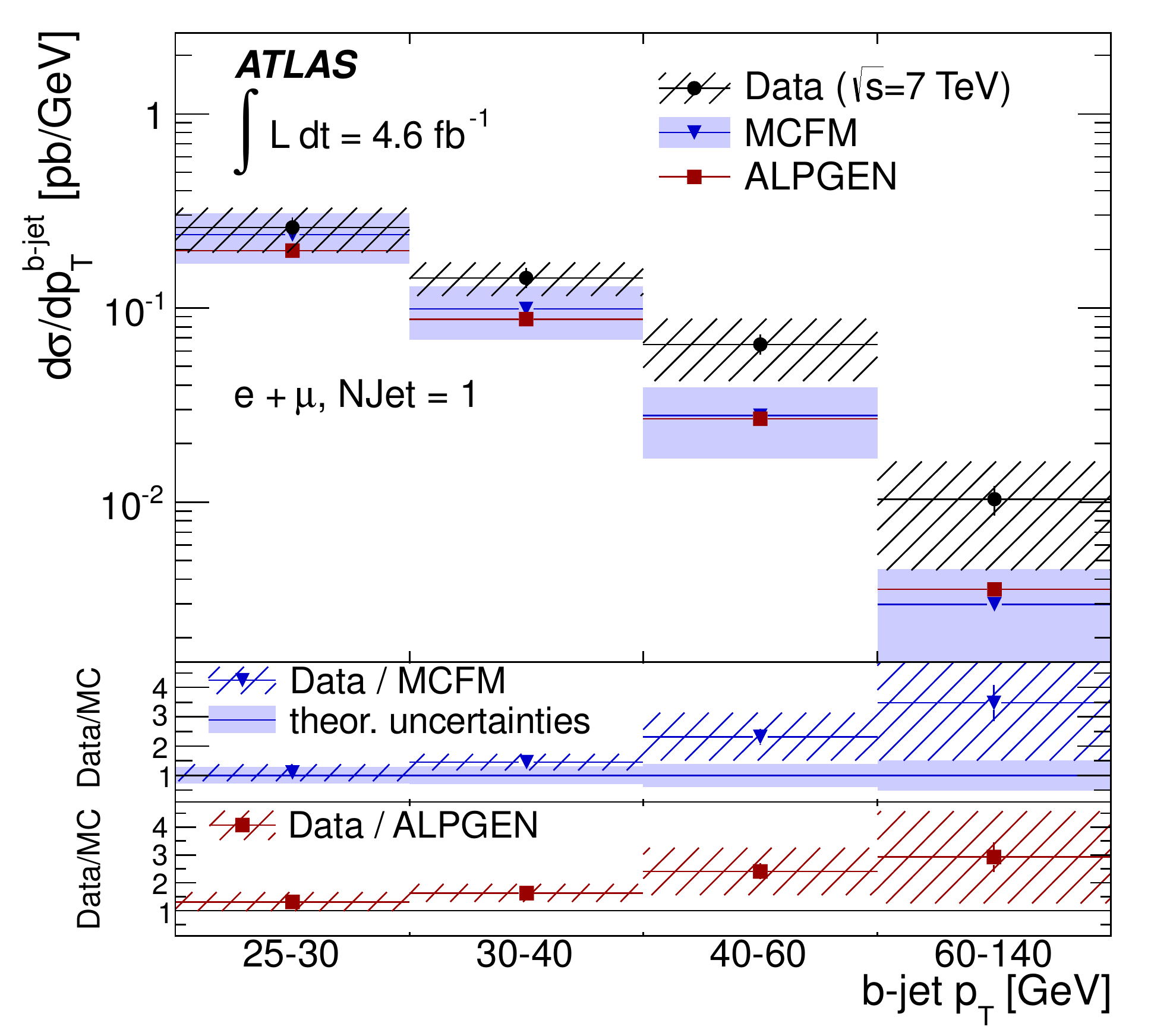}}
  \subfigure{\includegraphics[width=0.85\linewidth]{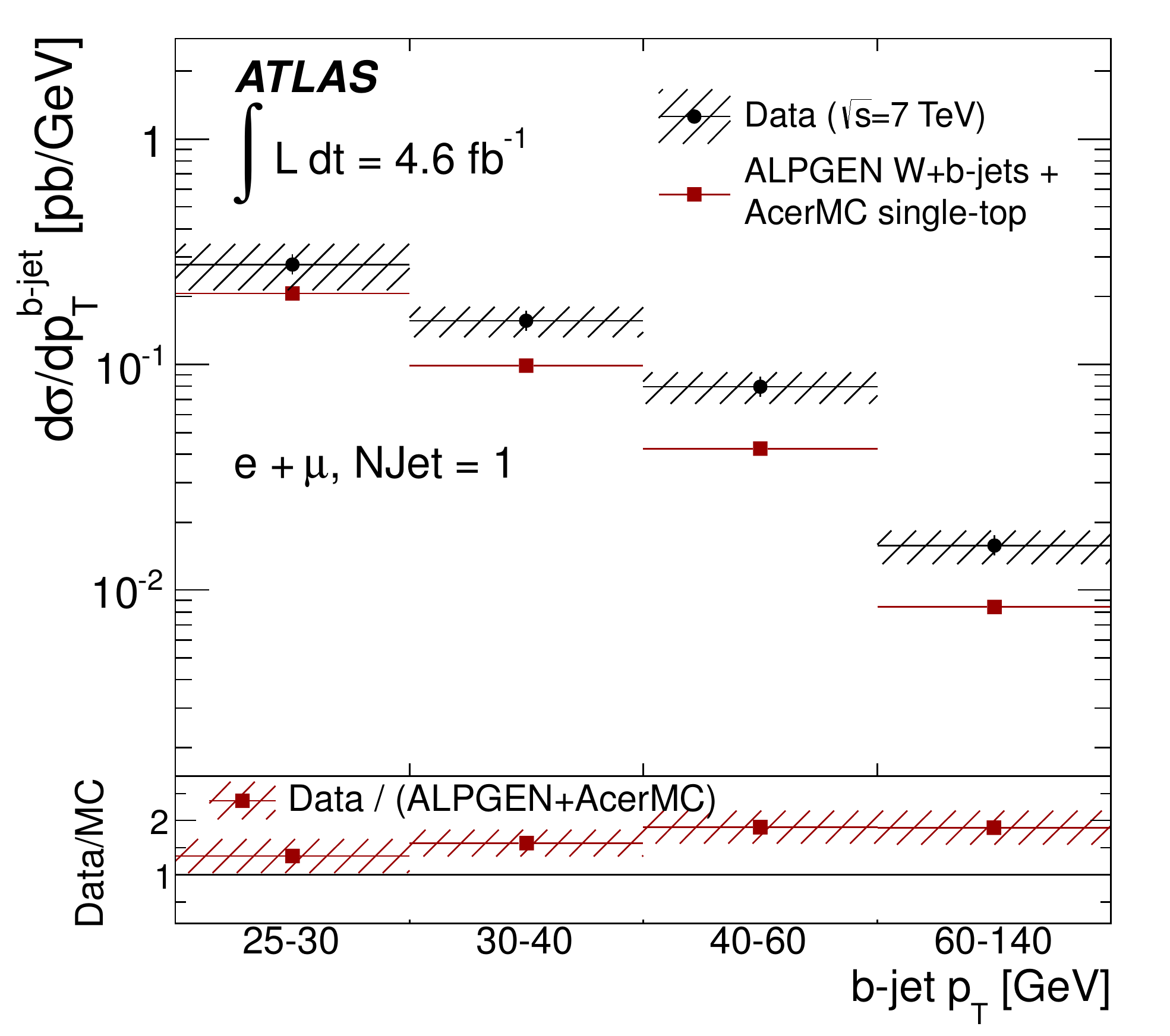}}
  \caption{Measured differential W+b-jets cross section as a function of the b-jet $\mathrm{ p_T }$~\cite{Aad:2013vka}, 
    in the 1-jet fiducial region, with (without) the single top subtraction is shown on top (bottom). 
    The statistical uncertainty is shown with inner error bar while statistical plus systematic is reported as outer error bar.
    The single top subtracted cross section is compared to NLO QCD predictions calculated with {\sc{Mcfm}} and to 
    {\sc{Alpgen}} interfaced to {\sc{Herwig}} and {\sc{Jimmy}} (scaled by the NNLO inclusive W normalization factor).
    The cross section measured without the single top subtraction is compared to {\sc{Alpgen}} interfaced to {\sc{Herwig}} and {\sc{Jimmy}} plus 
    {\sc{AcerMc}} interfaced with {\sc{Pythia}} and scaled to the NLO single-top cross-section.\label{fig:Wb2}} 
\end{figure}

The identification of $\mathrm{ W^\pm /Z }$ production using their leptonic decay 
modes~\footnote{Here the decay channel $\mathrm{ W\rightarrow \tau\nu }$ and $\mathrm{Z\rightarrow \tau\tau}$ are not considered.}
benefits from clean experimental signatures and small systematic uncertainties. Moreover, the theory calculation of the $\mathrm{W^\pm /Z}$ cross section is available up to 
NNLO in QCD~\cite{Gavin:2010az,Catani:2009sm,Anastasiou:2003ds} and NLO in EW, so these processes offer a solid benchmark for testing the predictions for the QCD corrections.

The inclusive cross section for $\mathrm{W^\pm/Z}$ production has been measured using the 2010 dataset~\cite{Aad:2011dm}. 
The cross section for $\mathrm{Z\rightarrow \ell\ell}$ as a function of $\mathrm{|y_Z|}$ and for 
$W^{\pm}\rightarrow \ell^{\pm}\nu$ as a function of $\mathrm{|\eta_{\ell^{\pm}}|}$ 
is shown in Fig.~\ref{fig:InclusiveWZ}, along with the theory predictions calculated at NNLO in QCD with 
{\sc{Fewz}} and {\sc{Dynnlo}} using various PDF sets. 
The uncertainty on the measurement is typically smaller than the spread among the theory predictions obtained 
with different PDF sets.
A global fit to the $\mathrm{W^\pm/Z}$ cross section
measurements and the HERA data is performed in order to extract the strange quark density of the proton~\cite{Aad:2012sb}; 
Fig.~\ref{fig:WZPDF} shows the result (``epWZ free'' fit) for the strange anti-quark PDF at $\mathrm{Q^2=1.9}$ GeV$^2$, compared with predictions from {\sc{Mstw2008}} 
(blue hatched) and {\sc{Abkm09}} (green hatched) PDF sets. The model used to fit the data is a NNLO QCD prediction with unconstrained $\mathrm{\overline s(x)}$.

The sensitivity to the strange quark density, driven by the $\mathrm{Z(\rightarrow \ell\ell)}$ cross section, is limited to the intermediate Bjorken-x range 
(from $\mathrm{x\sim 10^{-3}}$ up to $\mathrm{x\sim 10^{-1}}$). However, the x sensitivity reach can be extended measuring the Drell-Yan production in the high invariant mass region.
A measurement of the high mass Drell-Yan cross section has been reported in~\cite{ATLAS-CONF-2012-159}, for di-lepton invariant mass from $116$ up to $1500$ GeV. 
The data are well described by 
by a calculation at NNLO in QCD ({\sc{Fewz3.1}}) and NLO in EW interactions which includes also photon induced (PI) lepton pairs production, $\mathrm{\gamma\gamma\rightarrow\ell\ell}$, 
as shown in Fig.~\ref{fig:HighMassDY}.

Inclusive production of $\mathrm{Z/\gamma^*}$ allows resolution of initial state QCD radiation up to very soft scales. In particular, an ``optimal'' observable has been defined in 
Ref.~\cite{Banfi:2010cf}:
\begin{equation}
\mathrm{  \phi^{*}_{\eta}=\tan(\phi_{acop}/2)\sin(\theta^{*}_{\eta}) \, , }
\end{equation} 
where $\mathrm{ \phi_{acop} }$ is 
\begin{equation}
\mathrm{  \phi_{acop}=\pi-\Delta\phi(\ell^{+},\ell^{-})  }
\end{equation} 
and $\mathrm{ \sin(\theta^{*}_{\eta}) }$ is given by
\begin{equation}
\mathrm{  \sin(\theta^{*}_{\eta})=\sqrt{1-\tanh^2\left[\frac{(\eta(\ell^{-})-\eta(\ell^{+}))}{2} \right] } \, . }
\end{equation} 
The $\mathrm{ \phi^{*}_{\eta} }$ observable has been shown to be correlated to the Z transverse momentum,
but it has a better resolution at low transverse momentum with typical purity of $\mathrm{ \gtrsim 85\% }$.
The measurement of the differential $\mathrm{ \phi^{*}_{\eta} }$ spectrum has been presented in~\cite{Aad:2012wfa}; in Fig.~\ref{fig:PhiStar} it is displayed normalized to
the {\sc{ResBos}} prediction and is reported in Fig.~\ref{fig:PhiStar}. 
The ratio of the data to {\sc{ResBos}} are compared to two different predictions shown by a dashed line:  
a NLO+NLL QCD calculation from Ref.~\cite{Banfi:2012du} and a NNLO QCD prediction obtained with {\sc{Fewz2.1}}. 
The data are generally well described by the theory predictions and exhibit very high accuracy.

\section{Measurement of jet production in association with W/Z bosons\label{sec:WZJets}}

Jet production in association with a Z boson has been extensively studied in Ref.~\cite{Z+jets}.
The measured cross section for Z($\mathrm{ \rightarrow\ell\ell }$)+jets is shown in Fig.~\ref{fig:Z+jets} as a function of the inclusive jet multiplicity (N$\mathrm{_{jet}}$), 
the invariant mass of the two leading jets (m$\mathrm{^{jj}}$), and the azimuthal angle separation between the two leading jets 
($\mathrm{\Delta\phi^{jj}}$).
The data are compared to NLO multileg QCD predictions of {\sc{Blackhat+Sherpa}}~\cite{Frixione:2002ik}, and the {\sc{Alpgen}}, {\sc{Sherpa}}~\cite{Gleisberg:2008ta} 
and {\sc{Mc@nlo}}~\cite{Frixione:2002ik} 
event generators. 
The {\sc{Mc@nlo}} generator is unable to describe the data for high jet multiplicity as it is based on a NLO QCD calculation of the $\mathrm{2\rightarrow 2}$ matrix element; 
{\sc{Alpgen}} and {\sc{Sherpa}}, corrected for NNLO k-factor for inclusive Z cross section, provide a good description of the data; the NLO calculation of 
{\sc{Blackhat+Sherpa}}, shown without k-factor correction, describe well both the total cross section and the distributions profiles observed in the data.

Additional insight on the QCD radiation production can be achieved measuring observables free from the jet clustering algorithm. 
In Ref.~\cite{Aad:2013ueu} a measurement of the k$\mathrm{_t}$ splitting scales in $\mathrm{W\rightarrow \ell\nu}$ events has been presented. 
The k$\mathrm{_t}$ splitting scales are defined using the distance between constituents $\mathrm{i,j}$ and the beam $\mathrm{B}$ defined as:
\begin{equation}
\mathrm{  d_{i,j}=min(p^{2}_{Ti},p^{2}_{Tj})\,\frac{\Delta R_{ij}^{2}}{ R^{2} } \, , \quad d_{iB}=p_{Ti}^{2} \, ,   }
\end{equation} 
with $R=0.6$ being the radius parameter of the algorithm. At detector level the constituents are built from calorimeter clusters; 
stable particles are instead used at particle level.
The distribution of the hardest k$\mathrm{_t}$ splitting scale, $\mathrm{\sqrt{d_0}}$, in the $\mathrm{W\rightarrow \mu\nu}$ channel is shown in Fig.~\ref{fig:SplittingScales}.
The data are compared to the predictions from various MC generators, including {\sc{Mc@nlo}}, {\sc{Powheg}} {\sc{Alpgen}} and {\sc{Sherpa}}.

The copious production of W+2jets allows a test of the models of multiple parton interactions. 
Double-parton-interactions, DPI, in W+2jets events have been measured in~\cite{Aad:2013bjm} reporting the detector level fraction of DPI ($\mathrm{f_{DPI}^{(D)}}$) 
and the effective parameter area ($\sigma_{\mathit{eff}}$).
In Fig.~\ref{fig:DPI} the unfolded value of $\sigma_{\mathit{eff}}$ is compared to previous measurements preformed at different center of mass energy. 

Beyond the inclusive analysis of jet production in association with gauge boson, measurements involving heavy flavors drew attention recently 
since they are interesting channels to constrain the parton distribution functions and because it are background to $\mathrm{h\rightarrow b\overline b}$ observation in associated 
production with $\mathrm{W/Z}$ bosons.
The cross section for b-jets produced with a W boson has been recently measured using the 2011 dataset~\cite{Aad:2013vka}.
The cross section in jet multiplicity bins is reported in Fig.~\ref{fig:Wb1}. 
In Fig.~\ref{fig:Wb2} the b-jet p$\mathrm{_T}$ spectrum in the the 1-jet bin is shown; in the top plot at the top the signal is subtracted for the single top contribution while in 
the measurement shown at the bottom the single top is treated as signal.
The data are compared to NLO QCD ({\sc{Mcfm}}~\cite{Caola:2011pz}) as well as LO multileg predictions ({\sc{Alpgen}} and {\sc{Alpgen+AcerMC}}~\cite{Kersevan:2004yg}).

\section{Summary\label{sec:Conclusion}}
In this report the most recent QCD measurements performed with the ATLAS detector have been summarized. 

The production of jets as well as gauge bosons at the LHC were investigated in ATLAS with several measurements. 
The data are compared at particle level with the most advanced available theoretical predictions; 
a broad agreement between data and the predictions based on the Standard Model is found for wide spectrum of observables in various kinematical configurations. 
For several measurements the data were shown to be sensitive to NLO and NNLO QCD effects. In particular, electroweak boson cross section data tend to favor
NNLO and NLO, or LO multileg matrix element based predictions. 

\bibliography{apssamp}

\end{document}